\documentclass[apj]{emulateapj}

\usepackage{graphicx}
\usepackage{url}
\usepackage{ulem}

\usepackage{txfonts}
\usepackage{amssymb}
\usepackage{subfigure}
\usepackage{natbib}

\begin{document}

\newcommand{\HMS}[3]{$#1^{\mathrm{h}}#2^{\mathrm{m}}#3^{\mathrm{s}}$}
\newcommand{\DMS}[3]{$#1^\circ #2' #3''$}
\newcommand{\mrm}{\mathrm}
\newcommand{\degree}{$^\circ$}
\newcommand{\UNITS}[1]{\,\mathrm{#1}}

\newcommand{\hess}{H.E.S.S.}
\newcommand{\ngc}{NGC\,253}
\newcommand{\g}{$\gamma$}

\def\jcap{J. Cosmology Astropart. Phys.}%
          % Journal of Cosmology and Astroparticle Physics

\title{Spectral analysis and interpretation of the \g-ray emission
  from the Starburst galaxy \ngc\altaffilmark{*}}

\author{H.E.S.S. Collaboration,
A.~Abramowski \altaffilmark{1},
F.~Acero \altaffilmark{2},
F.~Aharonian \altaffilmark{3,4,5},
A.G.~Akhperjanian \altaffilmark{6,5},
G.~Anton \altaffilmark{7},
A.~Balzer \altaffilmark{7},
A.~Barnacka \altaffilmark{8,9},
Y.~Becherini \altaffilmark{10,11},
J.~Becker \altaffilmark{12}
% ...
}

\altaffiltext{1}{Universit\"at Hamburg, Institut f\"ur
  Experimentalphysik, Luruper Chaussee 149, D 22761 Hamburg, Germany}
\altaffiltext{2}{Laboratoire Univers et Particules de Montpellier,
  Universit\'e Montpellier 2, CNRS/IN2P3, CC 72, Place Eug\`ene
  Bataillon, F-34095 Montpellier Cedex 5, France}
\altaffiltext{3}{Max-Planck-Institut f\"ur Kernphysik, P.O. Box
  103980, D 69029 Heidelberg, Germany} \altaffiltext{4}{Dublin
  Institute for Advanced Studies, 31 Fitzwilliam Place, Dublin 2,
  Ireland} \altaffiltext{5}{National Academy of Sciences of the
  Republic of Armenia, Yerevan} \altaffiltext{6}{Yerevan Physics
  Institute, 2 Alikhanian Brothers St., 375036 Yerevan, Armenia}
\altaffiltext{7}{Universit\"at Erlangen-N\"urnberg, Physikalisches
  Institut, Erwin-Rommel-Str. 1, D 91058 Erlangen, Germany}
\altaffiltext{8}{Nicolaus Copernicus Astronomical Center, ul. Bartycka
  18, 00-716 Warsaw, Poland} \altaffiltext{9}{CEA Saclay, DSM/IRFU,
  F-91191 Gif-Sur-Yvette Cedex, France} \altaffiltext{10}{APC,
  AstroParticule et Cosmologie, Universit\'{e} Paris Diderot, CNRS/
  IN2P3,CEA/ lrfu, Observatoire de Paris, Sorbonne Paris Cit\'{e}, 10,
  rue Alice Domon et L\'{e}onie Duquet, 75205 Paris Cedex 13, France}
\altaffiltext{11}{Laboratoire Leprince-Ringuet, Ecole Polytechnique,
  CNRS/IN2P3, F-91128 Palaiseau, France} \altaffiltext{12}{Institut
  f\"ur Theoretische Physik, Lehrstuhl IV: Weltraum und Astrophysik,
  Ruhr-Universit\"at Bochum, D 44780 Bochum, Germany}
% ...

\email{stefan.ohm@le.ac.uk}
\email{denauroi@in2p3.fr}

\shortauthors{\hess\ Collaboration} 
\shorttitle{Analysis of the \g-ray emission from \ngc} 

\begin{abstract}
  Very-high-energy (VHE; $E\ge100\,\UNITS{GeV}$) and high-energy (HE;
  $100\,\UNITS{MeV} \leq E \leq 100\,\UNITS{GeV}$) data from \g-ray
  observations performed with the \hess\ telescope array and the {\it
    Fermi}-LAT instrument, respectively, are analysed in order to
  investigate the non-thermal processes in the starburst galaxy
  \ngc. The VHE \g-ray data can be described by a power law in energy
  with differential photon index $\Gamma=2.14 \pm 0.18_{\mathrm{stat}}
  \pm 0.30_{\mathrm{sys}}$ and differential flux normalisation at
  1\,TeV of $F_\mathrm{0}$ = $(9.6 \pm
  1.5_{\mrm{stat}}~(+5.7,-2.9)_\mrm{sys}) \times
  10^{-14}\UNITS{TeV^{-1}\,cm^{-2}\,s^{-1}}$. A power-law fit to the
  differential HE \g-ray spectrum reveals a photon index of
  $\Gamma=2.24 \pm 0.14_{\mathrm{stat}} \pm 0.03_{\mathrm{sys}}$ and
  an integral flux between 200\,MeV and 200\,GeV of
  $F(0.2-200\,\UNITS{GeV}) = (4.9 \pm 1.0_{\mrm{stat}} \pm
  0.3_{\mrm{sys}}) \times 10^{-9}\UNITS{cm^{-2}\,s^{-1}}$. No evidence
  for a spectral break or turnover is found over the dynamic range of
  both the LAT instrument and the \hess\ experiment: a combined fit of
  a power law to the HE and VHE \g-ray data results in a differential
  photon index $\Gamma=2.34 \pm 0.03$ with a p-value of 30\%. The
  \g-ray observations indicate that at least about 20\% of the energy
  of the cosmic rays capable of producing hadronic interactions is
  channeled into pion production. The smooth alignment between the
  spectra in the HE and VHE \g-ray domain suggests that the same
  transport processes dominate in the entire energy range. Advection
  is most likely responsible for charged particle removal from the
  starburst nucleus from GeV to multiple TeV energies. In a hadronic
  scenario for the \g-ray production, the single overall power-law
  spectrum observed would therefore correspond to the mean energy
  spectrum produced by the ensemble of cosmic-ray sources in the
  starburst region.
\end{abstract}


%%%%%%%%%%%%%%%%%%%%%%%%%%%%%%%%%%%%%%%%%%%%%%%%%
%
%    Section I : Introduction
%
%%%%%%%%%%%%%%%%%%%%%%%%%%%%%%%%%%%%%%%%%%%%%%%%%

\section{Introduction}
\label{sec:intro}

% Introduction to SB galaxies in general and connection to CRs
Starburst galaxies are galaxies that undergo an epoch of star
formation in a very localised region (the {\it starburst region}) at a
rate that is enhanced in comparison to other, so-called late-type
galaxies such as the Milky Way galaxy. It is believed that this
starburst activity is triggered either by galaxy mergers, a close
fly-by of galaxies, or by Galactic bar instabilities, where the
dynamical equilibrium of the interstellar gas gets disturbed. This
leads to the formation of regions of very high-density gas, usually at
the centre of the galaxy, and subsequently to star formation and a
strongly increased supernova (SN) explosion rate. SN remnant shocks
are widely believed to be acceleration sites of cosmic rays
(CRs). This is one reason why starburst regions might have a high CR
density. Given the high density of target material that is available
for p-p interactions and the production of $\pi^0$s, the starburst
nucleus is in addition a promising source of high-energy (HE;
$100\,\UNITS{MeV} \leq E \leq 100\,\UNITS{GeV}$) and very-high-energy
(VHE; $E\ge100\,\UNITS{GeV}$) \g\ rays. From energetic electrons also
Bremsstrahlung and Inverse Compton \g\ rays are expected. These
electrons may be either directly accelerated by the same processes as
the nuclear particles, or be generated in the decays of charged pions
from hadronic collisions. They might, however, also be produced in
different sources, like in pulsar wind nebulae. Starburst galaxies
have been predicted early on to be detectable by present \g-ray
instruments
\citep[e.g.][]{Voelk89,Voelk96,M82:Akyuz91,NGC253:Paglione96}.

% NGC 253
The spiral galaxy \ngc\ is the closest object in the southern sky that
belongs to the class of starburst galaxies. Its distance is measured
as $(2.6-3.9)$\,Mpc using different distance estimation techniques
\citep{NGC253:Davidge91,NGC253:Kara03,NGC253:Rekola05}. The reference
distance is $d = 2.6$\,Mpc \citep{NGC253:Davidge91} since this value
is used most widely in the literature to determine the properties of
\ngc. However, this reference distance has recently been convincingly
revised to 3.5 Mpc \citep{Dalcanton09}. The final numerical values
used below will therefore be needed to be scaled for consistency with
the revised distance value.

Compared to the Milky Way galaxy, \ngc\ exhibits an increased overall
star formation rate (SFR), with the SFR in the starburst nucleus being
comparable to that in the entire remaining disk of the galaxy. A SN
rate $\nu_\mathrm{SN}$ of this nucleus can be determined from the far
infrared (FIR) observations, since the FIR luminosity can be assumed
to be directly proportional to $\nu_\mathrm{SN}$
\citep{VanBuren94}. For \ngc\ as a whole the SN rate is estimated to
be $\approx 0.08\,\rm{yr^{-1}}$, with $\approx 0.03\,\rm{yr^{-1}}$ in
the starburst region \citep{NGC253:Engelbracht98}. By assessing the
SFR \citet{NGC253:Melo} found that it can amount to
$5\,M_\odot\,\rm{yr^{-1}}$ in the starburst nucleus alone which is
70\% of the SFR of the entire galaxy. The starburst region itself has
a cylindrical shape with a radius of $\approx 150$\,pc and a full
height of $\approx 60$\,pc perpendicular to the disk of the galaxy and
symmetric to its mid-plane with a volume $V_\mathrm{SB} \approx 1.2
\times 10^{62}\,(d/ 2.6\,\mathrm{Mpc})^3$\,cm$^3$
\citep{NGC253:Weaver02}.

% SFR in comparison with Milky Way
To understand the observed \g-ray emission, a simplified scenario is
considered in which the \g-ray production resulting from particle
acceleration in the part of the disk outside the starburst region is
neglected in comparison with that from the starburst region. The
reasons are the low average gas density and radiation field intensity
of the average Interstellar Medium, and the expected dominance of
energy-dependent diffusive particle losses from the disk --
quantitatively similar to the situation in the Milky Way galaxy. This
expectation is also consistent with the estimate of
\citet{MW:Strong10}, who find that the HE \g-ray luminosity of the
Milky Way galaxy is an order of magnitude lower than the \g-ray
luminosity of the starburst region of \ngc\
(cf. Section~\ref{sec:jointfit}).

The stellar winds from the early-type stars and the subsequent core
collapse SN explosions heat the lower-density parts of the surrounding
material, causing them to expand rapidly from the starburst region in
the form of a collective wind. The shocks from the SN explosions are
the primary accelerators of CRs in this scenario and their pressure
adds to the excess thermal gas pressure. The dense material in the
starburst region outside the SN remnants will remain essentially
non-ionised in this process and will not participate in the flow. The
$\pi^0$-producing CRs from the percolating wind flow are nevertheless
likely to penetrate also the dense gas in the starburst region which
therefore is a massive target for \g-ray production. The recent
detections of HE \citep{Fermi:NGC253M82} and VHE \g-ray emission from
the starburst galaxies \ngc\ \citep{HESS:NGC253} and M\,82
\citep{VERITAS:M82} appear to support this picture.

% Models for \g-ray emission and energy-loss processes
In a picture where there is quasi-steady equilibrium between
production and loss processes, the population of high-energy CRs
accelerated in \ngc's starburst nucleus is removed from the starburst
region predominantly via three different processes: (i) advective
removal of particles in the starburst wind; often also called a
``superwind'' \citep[see
e.g.][]{NGC253:Weaver02,NGC253:Zirakashvili06}, not to be confused
with the large-scale galactic ``disk wind'' \citep[see
e.g.][]{Breitschwerdt91,NGC253:Heesen09} that is primarily driven by
the general population of CRs and the hot gas, both produced in the
galactic disk, (ii) diffusion of particles from the source region, and
(iii) catastrophic inelastic (``p-p'') interactions. Energetic
electrons/positrons suffer, in addition, radiative losses. The
contributions of these components and the resulting \g-ray spectra
have been discussed by several groups \citep[see e.g.][]
{NGC253:Paglione96,
  HESS:NGC253_UL,NGC253:Domingo05,NGC253:Rephaeli10,NGC253:Lacki10,NGC253:Lacki11}
and are compared to the measurements presented here.

% Content
After the discovery of VHE \g-ray emission from \ngc\
\citep{HESS:NGC253}, we present for the first time a spectral analysis
of the VHE \g-ray data obtained by \hess\ in conjunction with the
analysis of a 30-month set of {\it Fermi-}LAT data, increased in size
by a factor of $\approx 3$ compared to the one used in the original
publication of the Fermi Collaboration on \ngc\
\citep{Fermi:NGC253M82}. These results are used to estimate the
properties of the underlying CR population, such as the particle
energy density, as well as to place constraints on CR transport and
the r.m.s. magnetic field strength within the starburst region.

%%%%%%%%%%%%%%%%%%%%%%%%%%%%%%%%%%%%%%%%%%%%%%%%%%%%%%%%%
%
%    Section II : H.E.S.S. Observations and Data Analysis
%
%%%%%%%%%%%%%%%%%%%%%%%%%%%%%%%%%%%%%%%%%%%%%%%%%%%%%%%%%
\section{\hess\ Observations and Data Analysis}
\label{sec:analysis}

%%% HESS general
\subsection{\hess\ instrument}
The High Energy Stereoscopic System (\hess) is an array of four
imaging atmospheric Cherenkov telescopes located in the Khomas
Highland of Namibia, 1800\,m above sea level. The telescopes are
identical in construction and each one comprises a 107\,m$^2$ optical
reflector composed of segmented spherical mirrors and a camera built
of 960 photomultiplier tubes. \hess\ utilises the {\it imaging
  atmospheric Cherenkov technique} \citep[see
e.g.][]{Hillas85}. Cherenkov light, emitted by the highly relativistic
charged particles in extensive air showers, is imaged by the mirrors
onto the camera. A single shower can be recorded by multiple
telescopes under different viewing angles, allowing stereoscopic
reconstruction of the primary particle direction and energy with an
average energy resolution of 15\% and an event-by-event spatial
resolution of 0.1\degree\ \citep{HESS:Crab}.

%%% Data set, Observations
\subsection{Data Set}
\ngc\ was observed with the \hess\ array in 2005 and from 2007 to 2009
for a total of 241~hours. After standard data quality selection, where
data taken under unstable weather conditions or with malfunctioning
hardware have been excluded, the total live time amounts to 177~hours
of three- and four-telescope observations that were used for the
generation of sky maps of the \g-ray emission and the reconstruction
of energy spectra. Observations were carried out at zenith angles of
1\degree~to 42\degree, with a mean value of 12\degree. Observations
have been performed in the {\it wobble-mode}, where the telescopes
were alternately pointed offset in RA and Dec from \ngc\ resulting in
an average pointing offset of 0.5\degree \citep{HESS:Crab}.

%%% Data analysis
\subsection{Data Analysis}
\label{sec:hess_analysis}
All results presented in the following were obtained using the {\it
  Model Analysis} \citep[{\it MA};][]{Model++} for event
reconstruction and background reduction and were cross-checked with
the boosted-decision-tree-based ({\it BDT}) Hillas parameter technique
described in detail in \citet{TMVA}. Two different sets of cuts were
used for {\it MA}: the {\it standard} cuts require a minimum shower
image intensity of 60\,p.e. in each camera. They maximise the
acceptance of \g-ray like events (at the expense of a larger
background) and are therefore used for energy spectra and sky maps;
the {\it faint} cuts, requiring a higher minimum intensity of
120\,p.e., resulting in improved angular resolution at the cost of
lower \g-ray acceptance, are used for position and extension
determination. Spectral results were derived using the {\it Reflected}
background model, whereas the {\it Ring} background model was utilised
to generate sky maps \citep{HESS:Background}. The analysis thresholds
for the {\it standard} and {\it faint} cuts configuration are 190\,GeV
and 250\,GeV for the {\it MA} method, respectively.

\begin{figure}
  \resizebox{\hsize}{!}{\includegraphics{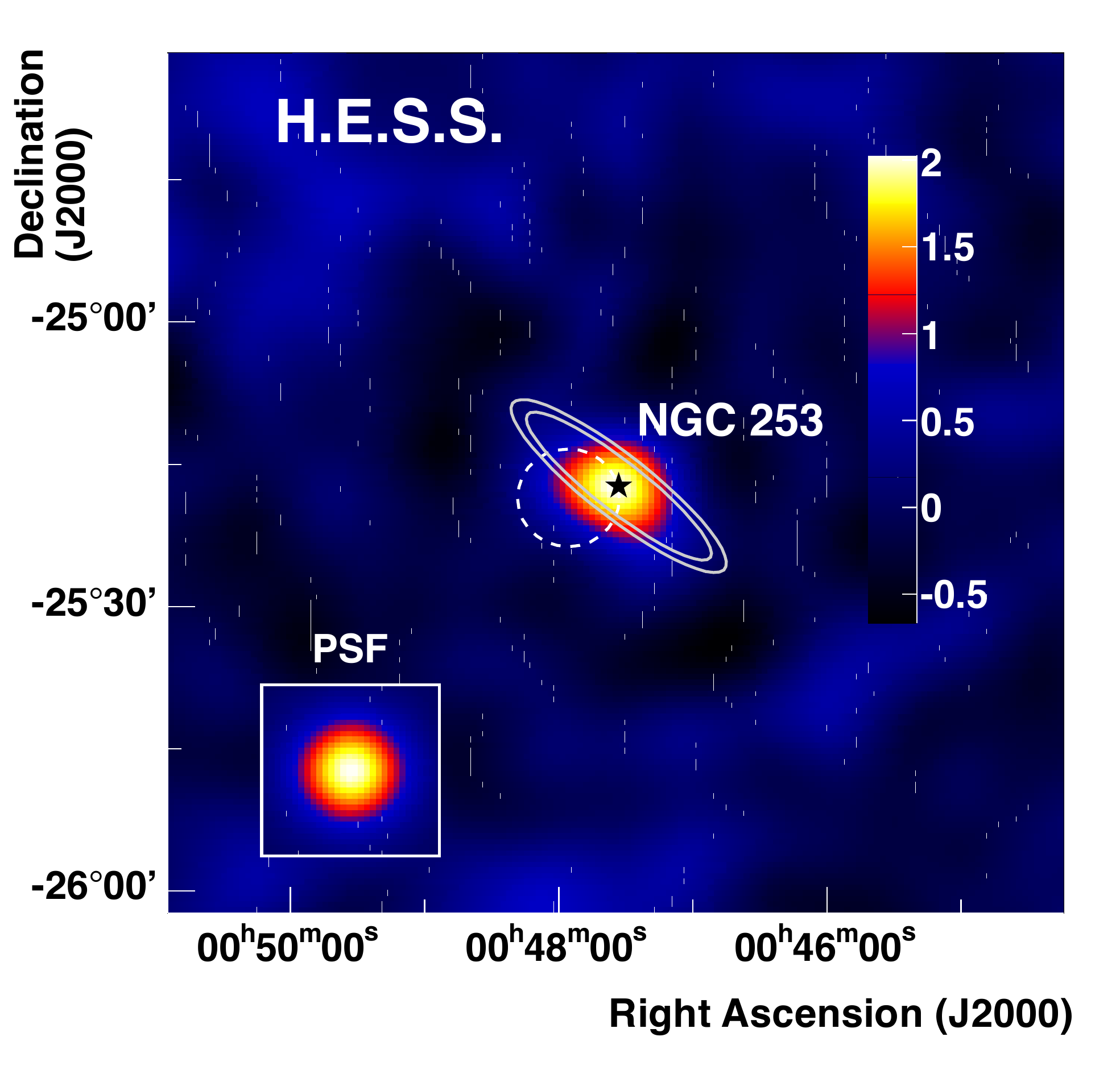}}
  \caption{Smoothed \hess\ \g-ray excess map in units of VHE \g-ray
    events per arcmin$^2$ of the $1.5^\circ\times 1.5^\circ$ FoV,
    centered on the position of \ngc. The image was smoothed with a
    Gaussian kernel of $3.9'$ r.m.s., the radius that corresponds to
    the PSF for this analysis. The black star marks the position of
    the optical centre of \ngc\ and the inlay represents the size of a
    point-like source as it would have been seen by \hess\ for this
    analysis. White contours depict the optical emission from the
    whole galaxy with contour levels of constant surface brightness of
    25 mag arcsec$^{-2}$ and 23.94 mag arcsec$^{-2}$ as used in
    \citet{NGC253:Pence80}. The dashed circle indicates the 95\%
      error contour of the best-fit position of the {\it Fermi}-LAT
      source (see also Table~\ref{tab:statstot}). (Colour version of
    this figure available online.)}
  \label{fig:ngc_excess}
\end{figure}

\subsection{Results}

A VHE \g-ray excess map for the $1.5^\circ\times 1.5^\circ$
Field-of-View (FoV) centred on the optical position of \ngc\ and
produced with {\it MA, standard} cuts, is shown in
Fig.~\ref{fig:ngc_excess}. The map has been smoothed with a 2D
Gaussian kernel of $3.9'$ r.m.s. to reduce the effect of statistical
fluctuations and matched to the PSF for this analysis. A total of $329
\pm 49$ excess events corresponding to a significance of $7.1\sigma$
\citep{LiMa} are found at the nominal position of \ngc. The overall
statistics of {\it MA} with both sets of cuts are shown in
Table~\ref{tab:ngc_data}. The best fit position of the source is RA
\HMS{00}{47}{34.3}$\pm 1.4^{\rm s}$, Dec \DMS{-25}{17}{22.6}$\pm 0.3'$
(J2000), compatible at the $<1\sigma$ level with the optical centre of
\ngc\ at RA \HMS{00}{47}{33.1} and Dec \DMS{-25}{17}{18} (J2000).

\begin{table}
  \begin{center}
    \caption{VHE \g-ray statistics of \ngc\ \label{tab:ngc_data}}
    \begin{tabular}{ccccccc}
      \tableline
      \tableline
      Cuts & $\theta^2_\mathrm{max}$ & $\mrm{N_{On}}$ & $\mrm{N_{Off}}$ & $1/\alpha$  & Excess & Significance\\
      & ($\deg^2$)& & & & & ($\sigma$) \\\hline
      {\it MA (standard)} & 0.01 & 2240 & 26224 & 13.72 & $329\pm 49$ & 7.1\\
      {\it MA (faint)} & 0.005 & 571 & 7816 & 20.13 & $183\pm 24$ & 8.4\\\tableline\tableline
    \end{tabular}
    \tablecomments{Number of \g-ray-like and background events along
      with the significance \citep{LiMa} and \g-ray excess as obtained
      for the \hess\ data using the {\it MA} method. $\alpha$ denotes
      the normalisation factor between signal and background
      exposure.}
  \end{center}
\end{table}

\begin{figure}
  \centering
  \resizebox{\hsize}{!}{\includegraphics{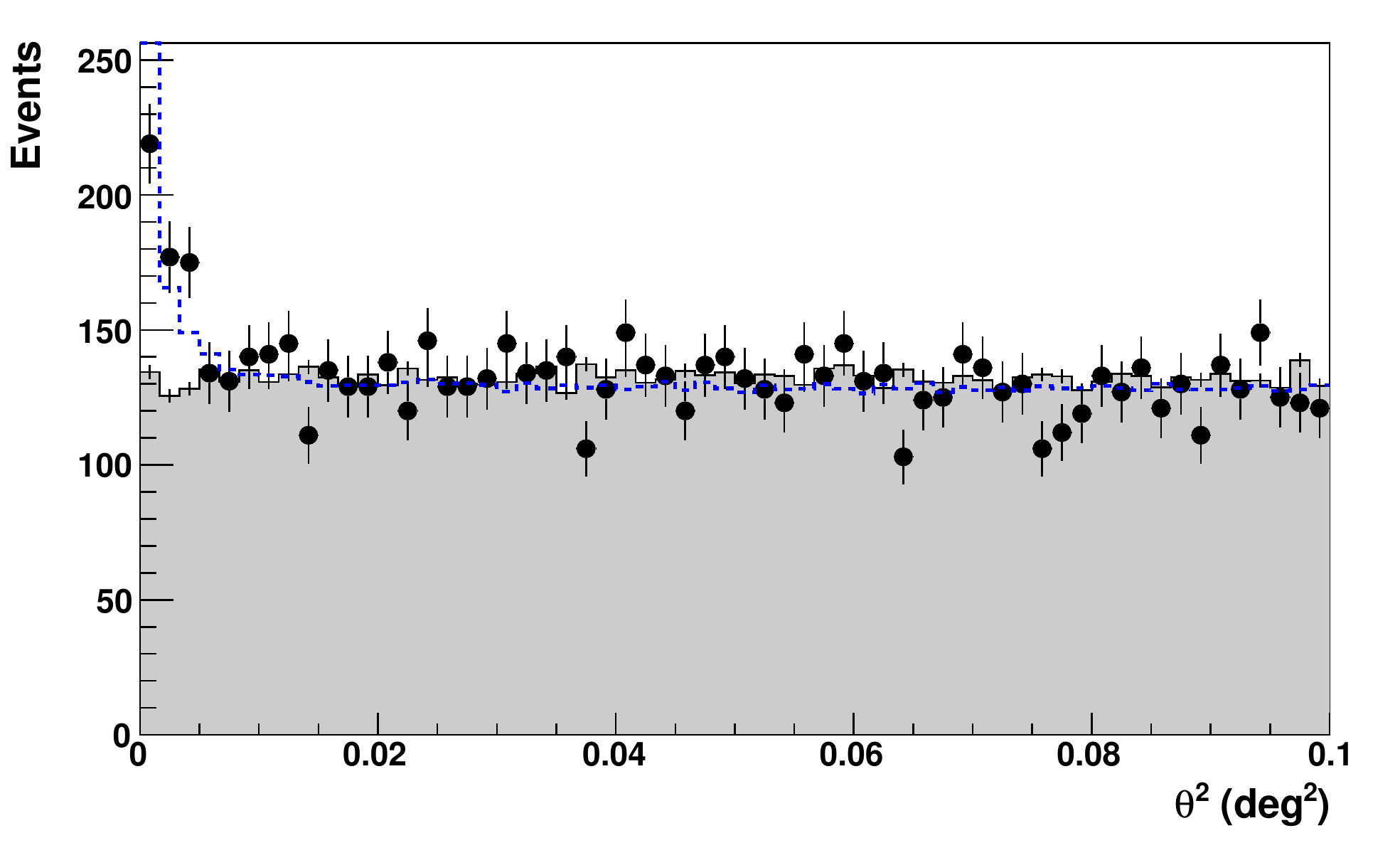}}

  \caption{Distribution of ON events around \ngc\ and of OFF events
    from background control regions as obtained with {\it MA, faint}
    source cuts. The squared angular distribution has been produced
    using the {\it Reflected} background model. Also shown is the
    point spread function of the instrument for this analysis as
    dotted line, assuming the \g-ray emission originates from the
    optical centre of \ngc. The normalisation of the model is adjusted
    to match the total \g-ray excess in the range
    0\,deg$^2-0.01$\,deg$^2$.}
  \label{fig:ngc_thetasq}
\end{figure}

The squared angular distribution of \g-ray candidate and background
events relative to the position of \ngc\ as shown in
Fig.~\ref{fig:ngc_thetasq} is consistent with point-like
emission. This constrains a potential source extension to less than
$2.4'$ at $3\sigma$ confidence level \citep[cf. the extent of the
starburst region in $^{12}$CO(2-1) is about $\approx 0.4'\times
1.0'$;][]{NGC253:Sakamoto11}. In the standard picture, where the HE
and VHE \g-ray emission from starburst galaxies originates from
diffuse CR interactions, no variability of the \g-ray signal is
expected. The yearly light curve of the \g-ray emission is stable over
the four years of observations within errors. A fit of a constant flux
to the yearly light curve yields a mean integral flux above 1\,TeV of
$(8.5\pm1.8)\times 10^{-14}\,\UNITS{cm^{-2}}\,\UNITS{s^{-1}}$ with a
$\chi^2$ of 3.3 for 3 degrees of freedom and is stable over the four
years of observations within errors.

The \hess\ data set previously published in \citet{HESS:NGC253}
comprises 119\,hrs of good quality data, collected in the years 2005,
2008 and 2009, and represents 2/3 of the data set used in this
work. Based on this larger data set, a better determination of the
spectral characteristics is now possible. All results presented in
that publication are consistent with the findings presented here.

\subsection{Spectrum}
\label{sec:hess_spec}
\begin{figure*}[t]
  \centering
  \resizebox{0.7\hsize}{!}{\includegraphics{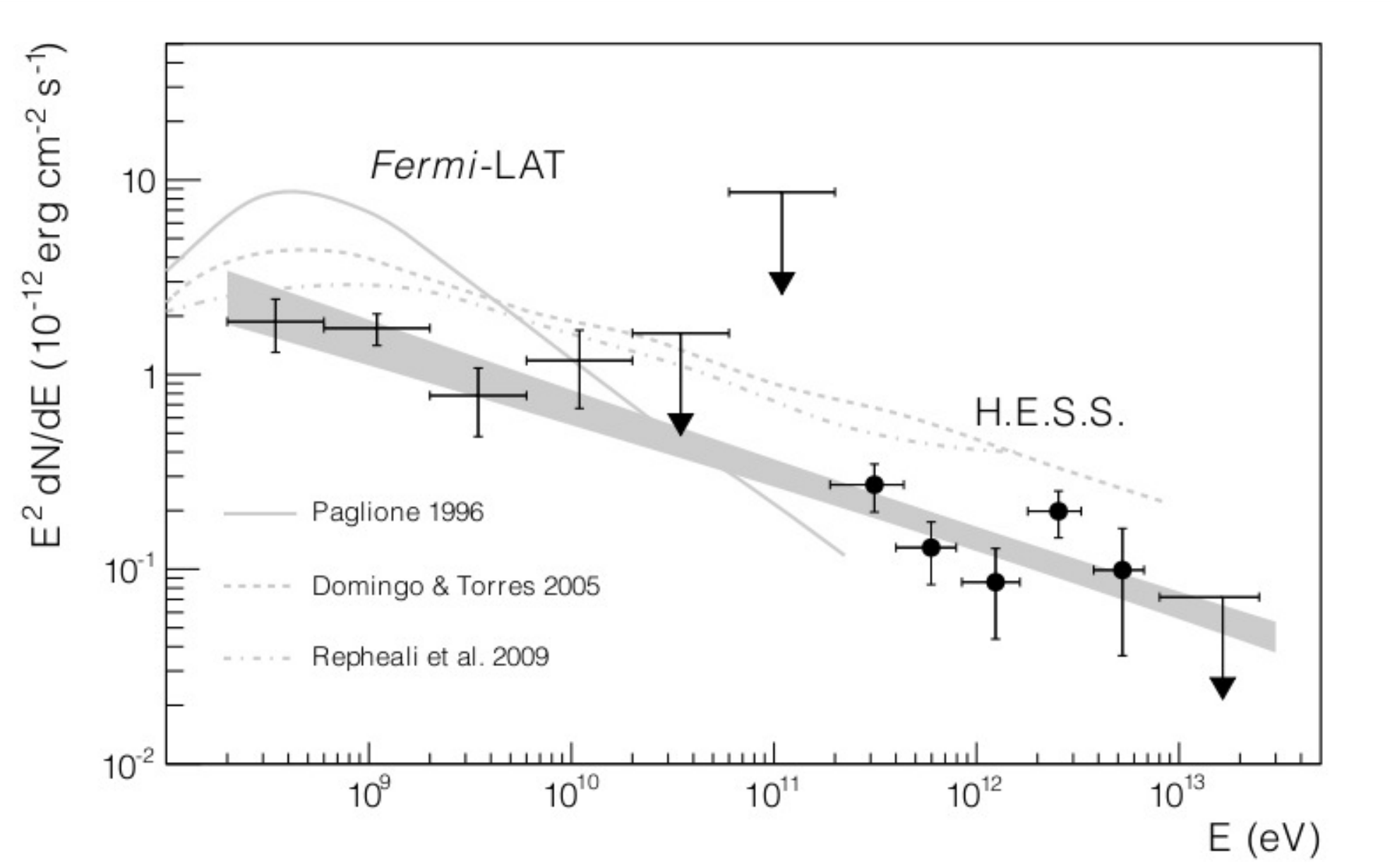}}
  \caption{Differential \hess\ energy spectrum of \ngc\ as obtained
    with {\it MA} (shown as circles). Also shown is the LAT energy
    spectrum as obtained in the analysis of the 30 months of data as
    described in the text (shown as crosses). For both spectra
    $1\sigma$ error bars are shown for spectral points and 95\% upper
    limits according to \citet{FeldmanCousins98}. The shaded area
    represents the $1\sigma$ confidence band from the simultaneous fit
    to the {\it Fermi}-LAT and H.E.S.S. data. Also shown are the
    predictions from \citet{NGC253:Paglione95} (solid grey),
    \citet{NGC253:Domingo05} (dashed grey) and
    \citet{NGC253:Rephaeli10} (dash-dotted grey).}
\label{fig:ngc_spectrum}
\end{figure*}

\begin{table*}
  \begin{center}
    \caption{\hess\ spectral results of \ngc. \label{tab:ngc_spectrum}}
    \begin{tabular}{lcccc}
      \tableline\tableline
      Data
      &$E_{\UNITS{th}}$
      &$F_0$
      &$F(>E_{\UNITS{th}})$
      &$\Gamma$
      \\      
      {} 
      & (GeV)
      & ($\UNITS{TeV^{-1}\,cm^{-2}\,s^{-1}}$)
      & ($\UNITS{cm^{-2}\,s^{-1}}$)
      &  \\
      \tableline
      this analysis by {\it MA} & 190 & $(9.6 \pm 1.5)\times10^{-14}$ & $(5.6 \pm 1.2)\times10^{-13}$ & $2.14 \pm 0.18$ \\
      prev. analysis \citep{HESS:NGC253} & 220 & & $(5.5 \pm 1.0)\times10^{-13}$   &  \\\tableline\tableline
    \end{tabular}
    \tablecomments{VHE \g-ray spectral results as shown in
      Fig.~\ref{fig:ngc_spectrum} for the {\it MA}. The photon index
      $\Gamma$ is derived from a fit of a power law to the
      spectrum. Only statistical errors are given. A comparison with
      the integral flux given in \citet{HESS:NGC253} is also
      provided.}
  \end{center}
\end{table*}

The differential energy spectrum derived from this data set is shown
in Fig.~\ref{fig:ngc_spectrum} and Table~\ref{tab:ngc_spectrum} and is
well described by a power law:
$\mathrm{d}N/\mathrm{d}E=F_0\cdot(E/1\,\mathrm{TeV})^{-\Gamma}$ with
photon index $\Gamma = 2.14 \pm 0.18_{\mathrm{stat}} \pm
0.30_{\mathrm{sys}}$ and flux normalisation $F_0 = (9.6 \pm
1.5_{\mrm{stat}} ~(+5.7,-2.9)_\mrm{sys}) \times
10^{-14}\UNITS{TeV^{-1}\,cm^{-2}\,s^{-1}}$ with a chance probability
of $7\%$.  This yields an integral flux above the energy threshold of
190\,GeV of $F(>190\UNITS{GeV})=(5.6 \pm 1.2) \times
10^{-13}\UNITS{cm^{-2}\,s^{-1}}$. With a flux of $(0.21\pm 0.05)$\% of
the Crab nebula flux above the energy threshold, \ngc\ is the source
with the lowest VHE \g-ray flux detected so far.

Since the VHE \g-ray signal from \ngc\ is so weak, systematic effects
could potentially influence the spectral reconstruction. In-depth
systematic checks on the background subtraction and the spectrum
calculation have been performed with the {\it MA} and {\it BDT}
method. These tests have been used to estimate the systematic
uncertainty of the VHE \g-ray spectrum presented here. For instance,
the difference in the data selection procedures leads to a difference
in live time of 12\% and hence to a small difference in the data sets
used to reconstruct the spectrum. Furthermore, the Earth's magnetic
field bends charged particles in EASs and influences their
development. This affects observables such as the shower image
orientation in the Cherenkov camera, the image shape and hence the
stereoscopic reconstruction. Although this effect is expected to be
small, it is taken into account in the reconstruction of the energy
spectrum of \ngc. In order to test the background systematics, harder
selection cuts on \g-ray like events have been performed, e.g. the
{\it faint} cuts introduced in Section~\ref{sec:hess_analysis} and
alternate background estimation techniques (such as the {\it template
  background}) have been studied as well. Precise comparison between
the actual level of background in the whole field of view and
predictions from the background model, excluding a circle of
0.25\degree around \ngc, indicate that the background level is
controlled at a level of $\approx 1.5\%$ over the full field of view,
with the magnetic field introducing an azimutal asymmetry of $\pm
2\%$.  The magnetic field as well as potential systematic effects in
the background determination mainly affect low-energy
events. Therefore, additional tests have been performed, where only
events with reconstructed energies above 0.6\,TeV have been used in
the spectral fit. All systematic tests have been performed with both
analysis chains and resulted in a spread in normalisation between the
two analyses of $+60\%, -30\%$ and a variation in differential photon
index of $\Delta\Gamma_{\UNITS sys}\leq0.3$.

%%%%%%%%%%%%%%%%%%%%%%%%%%%%%%%%%%%%%%%%%%%%%%%%%%%%%%%%%
%
%    Section III : Fermi Observations and Data Analysis
%
%%%%%%%%%%%%%%%%%%%%%%%%%%%%%%%%%%%%%%%%%%%%%%%%%%%%%%%%%
\section{{\it Fermi}-LAT Observations and Data Analysis}
\label{sec:fermi}

Based on eleven months of data, \ngc\ has also been detected in the HE
\g-ray regime by the {\it Fermi}-LAT instrument
\citep{Fermi:NGC253M82}. It was later confirmed as a HE \g-ray emitter
based on 12 and 24 months of data \citep{Fermi:1yr,Fermi:2yr}. The
analysis and the results of a larger 30 month data set are presented
in the following.

\subsection{{\it Fermi}-LAT instrument}
The {\it Fermi}-LAT instrument is a pair-conversion telescope, capable
of detecting \g\ rays in the energy range between 20\,MeV and
300\,GeV. It consists of a tracker for the reconstruction of the
particle direction, a calorimeter which measures the energy of the
incident particle, and an anti-coincidence system designed to suppress
the charged-particle background. Data are normally recorded in survey
mode, in which the whole sky is covered every two orbits. The
instrument FoV is $\approx 2.4\,\UNITS{sr}$ and it provides an angular
resolution of $<1^\circ$ at 1\,GeV and $<0.2^\circ$ at 10\,GeV. A full
description of the mission- and instrument-related details can be
found in \citet{Fermi:Inst}.

\subsection{Data Set and Data Analysis}

The data set presented in the following comprises data from the
commissioning of Fermi on 2008 August $4^{\mathrm{th}}$, (MJD 54682)
until 2011 February $3^{\mathrm{rd}}$ (MJD 55595). The data analysis
has been performed using events with reconstructed energies between
200\,MeV and 200\,GeV and utilising the Fermi Science Tools (FST)
package, version
v9r18p6\footnote{\url{http://fermi.gsfc.nasa.gov/ssc/data/analysis/documentation/Cicerone/}}.
Events of the ``Diffuse'' source class have been analysed using the
{\it P6\_V3\_Diffuse} instrument response functions
(IRFs). Additionally, events with zenith angles $> 105^{\circ}$ were
excluded from the analysis due to a significant contribution of
Earth-limb \g\ rays\footnote{Note that as a cross check, the data have
  also been analysed with the more recent FST release, version v9r23p1
  and the {\it P7Source\_V6} IRFs, and found to give consistent
  results.}.

\begin{table*}
  \begin{center}
    \caption{{\it Fermi}-LAT spectrum and position of \ngc. \label{tab:statstot}}
    \begin{tabular}{c c c r@{$\:\pm\:$}l r@{$\:\pm\:$}l c}
      \tableline\tableline
      RA (J2000)
      & Dec (J2000)
      & $r_{95}$
      & \multicolumn{2}{c}{$F(0.2-200\,\mathrm{GeV})$}
      & \multicolumn{2}{c}{$\Gamma$}
      & TS 
      \\
      (deg) 
      & (deg) 
      & (deg)
      & \multicolumn{2}{c}{($10^{-9}$\,cm$^{-2}$\,s$^{-1}$)} 
      & \multicolumn{2}{c}{}
      & \\
      \tableline
      11.982 & -25.309 & 0.095 & 4.9 & $1.0_{\rm stat} \pm 0.3_{\rm sys}$ & 2.24 & $0.14_{\rm stat} \pm 0.03_{\rm sys}$ & 105 \\
      \tableline\tableline
    \end{tabular}
    \tablecomments{Best fit position and results of the maximum
      likelihood analysis between 200\,MeV and 200\,GeV.}
  \end{center}
\end{table*}

The best fit position of \ngc\ has been determined by means of a
maximum likelihood method, using the FST tool $gtfindsrc$. All events
in a Region-of-Interest (RoI) of $10^\circ$ around this best fit
position have been used, and all sources within $15^\circ$ were
modeled to produce energy spectra and to calculate the test statistic
(TS)\footnote{$\sqrt{\mathrm{TS}}$ is a measure of the significance
  and defined as $\mathrm{TS} = -2\Delta$log(likelihood) between
  models including and excluding the source \citep{Mattox1996}.} of
the source. At the best fit position at RA \HMS{00}{47}{55.7} and Dec
\DMS{-25}{18}{32.4} (J2000) ($r_{95} = 5.7'$, 95\% confidence level) a
TS value of 105, corresponding to a statistical significance of
$\approx 10\,\sigma$ is found.

Using the FST $gtlike$ tool and an un-binned maximum likelihood
fitting procedure, the energy spectrum of \ngc\ has been derived. For
this purpose, all sources listed in the {\it Fermi}-LAT 1-year
catalogue in the $15^\circ$ region around the best fit position were
modeled with a power law ($\mathrm{d}N/\mathrm{d}E \propto
E^{-\Gamma}$), with differential photon index $\Gamma$. In a second
step, residual sources in the TS map in the RoI with a
$\mathrm{TS}>16$ were included in the model as well. Compared to the
1-year and 2-year catalogue 20 and 16 more candidates, respectively,
are included in the source model used in this work. In the
minimisation procedure, the integral flux in the energy range of
interest and the photon index were left as free parameters for all
sources in the RoI and fixed for all sources between 10 and
15$^\circ$. The diffuse Galactic and Extragalactic background
components were modeled with the files $gll\_iem\_v02.fit$ and
$isotropic\_iem\_v02.txt$\footnote{\url{http://fermi.gsfc.nasa.gov/ssc/data/access/lat/BackgroundModels.html}},
respectively. The energy spectrum is best described by a power law in
energy with $\Gamma = 2.24 \pm 0.14_{\mathrm{stat}} \pm
0.03_{\mathrm{sys}}$ and an integral flux of $F(0.2-200\,\UNITS{GeV})
= (4.9 \pm 1.0_{\mrm{stat}} \pm 0.3_{\mrm{sys}}) \times
10^{-9}\UNITS{cm^{-2}\,s^{-1}}$. The positional and spectral results
of the likelihood analysis are summarised in Tab.~\ref{tab:statstot}.

The flux in each energy band as summarised in
Table~\ref{tab:statsbins} and shown in Fig.~\ref{fig:ngc_spectrum} has
been reconstructed in the same way as for the full energy range, where
the photon index and integral flux in that band were left as free
parameters again. Note that all spectral points agree within
$1\,\sigma$ statistical error with the spectral points and upper
limits as reported in \citet{Fermi:NGC253M82}.

The systematic uncertainty on the total spectrum has been estimated
following the bracketing method \citep[e.g.][]{Fermi:Blazars09}, where
the effective area is shifted up- and downward according to its
systematic uncertainty in the corresponding energy range. Following
\citet{Fermi:1yr}, the systematic uncertainty on the flux as inferred
by the systematic uncertainty of the effective area is of the order of
10\% at 100\,MeV, 5\% at 0.5\,GeV and 20\% at 10\,GeV. As for the VHE
\g-ray light curve, there is no significant variability seen in the
yearly HE \g-ray emission.

\begin{table}
 \begin{center}
   \caption{{\it Fermi}-LAT spectral results of \ngc. \label{tab:statsbins}}
   \begin{tabular}{l r@{$\:\pm\:$}l l}
     \tableline\tableline
     $E_\mathrm{min} - E_\mathrm{max}$ 
     & \multicolumn{2}{c}{$F_{\rm{NGC\,253}}$}
     & syst. uncer. 
     \\
     (GeV)
     & \multicolumn{2}{c}{($10^{-12}$\,erg\,cm$^{-2}$\,s$^{-1}$)}
     & (\%) \\
     \tableline
     $0.2-0.6$ & 1.9 & 0.6 & 10  \\
     $0.6-2.0$ & 1.7 & 0.3 & 10  \\
     $2.0-6.0$ & 0.8 & 0.3 & 15  \\
     $6.0-20.0$ & 1.2 & 0.5 & 20  \\
     $20.0-60.0$ &  \multicolumn{2}{l}{$<1.7$} & 20  \\
     $60.0-200.0$ & \multicolumn{2}{l}{$<8.7$} & 20  \\
%     $0.2-0.6$ & 1.87 & 0.57 & 10  \\
%     $0.6-2.0$ & 1.73 & 0.32 & 10  \\
%     $2.0-6.0$ & 0.78 & 0.30 & 15  \\
%     $6.0-20.0$ & 1.18 & 0.51 & 20  \\
%     $20.0-60.0$ &  \multicolumn{2}{l}{$<1.63$} & 20  \\
%     $60.0-200.0$ & \multicolumn{2}{l}{$<8.65$} & 20  \\
     \tableline\tableline
   \end{tabular}
   \tablecomments{HE \g-ray spectral results of the maximum likelihood
     analysis in energy bands as shown in
     Fig.~\ref{fig:ngc_spectrum}. $1\sigma$ statistical errors and
     95\% upper limits are given. Systematic uncertainties have been
     obtained using the bracketing method \citep[see
     e.g.][]{Fermi:Blazars09}.}
 \end{center}
\end{table}

\section{The HE-VHE \g-ray spectrum}
\label{sec:jointfit}

Within the $1\sigma$ statistical errors, the \hess\ measurement and
the Fermi results are compatible, in respective value of photon index
as well as respective normalisation. A simultaneous, single power law
fit to the \hess\ and {\it Fermi}-LAT spectral points results in a
photon index of $\Gamma_{\UNITS{sim}} = 2.34\pm0.03$ and an energy
flux at 1\,GeV of $(1.5 \pm 0.2)\times
10^{-12}$\,erg\,cm$^{-2}$\,s$^{-1}$ \footnote{Note that no
  forward-folding technique has been applied and the fit is performed
  only taking into account statistical errors.}. This corresponds to
an integral energy flux above 0.2\,GeV of
$F_\mathrm{\gamma}^\mathrm{meas}(> 0.2\,\mathrm{GeV}) \simeq 5.3 \times
10^{-12}$\,erg\,cm$^{-2}$\,s$^{-1}$. The fit has a $\chi^2$ of 8.27
for 7 degrees of freedom and a p-value of 30\%. Note that the somewhat
steeper index for the simultaneous fit compared to the individual
measurements is fully compatible within the 1$\sigma$ statistical
uncertainties of the {\it Fermi} and \hess\ photon index. Even when
taking into account a 30\% downward shift to account for the
systematic error in reconstructed normalisation of the VHE \g-ray
spectrum as presented in Section~\ref{sec:hess_spec}, the single power
law fit has a p-value of 8.8\%.

The fit of a broken power law to the HE and VHE \g-ray points with a
break energy between 10 and 300\, GeV (between \hess\ and {\it Fermi})
results in a best-fit index change of $\Delta\Gamma = 0.1 \pm 0.3$
with break energy at 300\, GeV. Although, a photon index change
between the {\it Fermi} and \hess\ energy range can not be excluded,
the hypothesis of no index change is favoured given the available
data. This result suggests that in order to explain the available data
no spectral model which includes a break or turnover is needed.

The integral (HE-VHE) flux above 200\,MeV corresponds to a \g-ray
luminosity of $L_\mathrm{\gamma}^E(> 200\,\mathrm{MeV}) \simeq 7.8
\times 10^{39}$\,erg\,s$^{-1}$, for a distance of 3.5\,Mpc. This is
about a factor of ten larger than the \g-ray luminosity
$L_\mathrm{\gamma}^{\mathrm{MW}}(> 200\, \mathrm{MeV}) \simeq (7-10)
\times 10^{38}$\,erg\,s$^{-1}$ for the Milky Way galaxy, as estimated
by \citet{MW:Strong10}.

%%%%%%%%%%%%%%%%%%%%%%%%%%%%%%%%%%%%%%%%%%%%%%%%%%%%%%%%%
%
%    Section IV : Discussion section 
%
%%%%%%%%%%%%%%%%%%%%%%%%%%%%%%%%%%%%%%%%%%%%%%%%%%%%%%%%%
\section{Discussion}

The results presented in the previous sections have a number of
interesting implications for the nature and properties of the \g-ray
source in the starburst nucleus of \ngc. Note that the discussion is
based on the observationally favoured result that the spectrum of
\ngc\ can be described by a single power law from the HE to the VHE
\g-ray regime. Quantities used in the following are summarised in
Table~\ref{tab:quant}. This section is organised as follows: firstly
the aspect of dominance of a hadronic scenario is discussed. Secondly,
arguments for an energy-independent particle transport in the
starburst region are given, the consequences of this picture are
explored, and limits on the diffusion coefficient are
presented. Thirdly, the \g-ray flux estimated from the supernova rate
is compared to the experimental results. Fourthly, the magnetic field
in the starburst region is estimated and finally, the contribution of
discrete sources to the \g-ray luminosity of \ngc\ is assessed.

\begin{table}
  \begin{center}
    \caption{Symbols, units and descriptions of quantities used in
      the discussion \label{tab:quant}}
    \begin{tabular}{ccp{4.5cm}}
      \tableline
      \tableline
      Symbol & Units & Description\\\hline
      $\epsilon$ & GeV & energy of individual particle \\
      $E_\mathrm{pp}^{\pi}$ & GeV & total non-thermal energy of
      pion-producing particles\\
      $\tilde{Q}$ & GeV$^{-1}$\,cm$^{-3}$\,s$^{-1}$ & differential
      volumetric particle source term\\
      $Q$ & GeV\,s$^{-1}$ & volume-integrated particle source term\\
      $Q^{\pi}$ & GeV\,s$^{-1}$ & volume-integrated source term of
      pion-producing particles \\
      \tableline\tableline
    \end{tabular}
  \end{center}
\end{table}

\subsection{Dominance of hadronic \g-ray emission} Early model
predictions for the various contributions to the overall HE and VHE
\g-ray spectral energy distribution \citep[e.g.][]{NGC253:Paglione96}
and subsequent more detailed calculations
\citep[e.g.][]{NGC253:Domingo05,NGC253:Rephaeli10,NGC253:Lacki10,NGC253:Lacki11},
all consider {\it diffuse \g-ray emission} and neglect the
contributions of discrete sources. They are roughly compatible with
the observational results. Three such model curves are shown in
Fig.~\ref{fig:ngc_spectrum}.

These largely phenomenological models use primary electron-to-proton
ratios similar to those inferred for the Interstellar Medium (ISM) of
the Milky Way galaxy and parameterise the diffusive escape times from
the starburst region, while assuming advection speeds of the order of
several hundred km\,s$^{-1}$. Also the secondary leptonic component
from the decay of charged pions is included in the calculation of the
expected Bremsstrahlung and Inverse Compton \g-ray emission.
In detail the three model spectra in Fig.~\ref{fig:ngc_spectrum} are
based on very different assumptions regarding the relative magnitudes
of the advective ($\tau_\mathrm{ad}$) and energy-dependent diffusive
($\tau_\mathrm{diff}$) particle escape times on the one hand, and on
the assumed source spectra on the other. These models assume
$\tau_\mathrm{diff} = \tau_0 (\epsilon_\mathrm{p}/(1 \mathrm{GeV})^{-0.5}$,
where $\epsilon_\mathrm{p}$ is the proton energy. In
\citet{NGC253:Paglione96} $\tau_\mathrm{ad} > \tau_\mathrm{diff}$ for
all energies observed\footnote{Due to a misprint in that paper the
  $\tau_0$ in their Fig.~6 is in fact 1 Myr instead of the quoted
  $\tau_0 = 10$\,Myr (T.A.D. Paglione, private communication).}. With
an assumed proton differential source spectral index $s=2.2$ the
\g-ray spectral energy density (SED) has a rather steep dependence
$\propto \epsilon_\gamma^{-0.7}$ on the \g-ray energy $\epsilon_\gamma$ due to the
dominance of diffusive escape. \citet{NGC253:Rephaeli10} obtain
$\tau_\mathrm{ad} > \tau_\mathrm{diff}$. However, their source
spectrum is assumed to be quite hard ($s=2.0$). These authors model
the emission from the galactic disk as a whole and find approximately
a SED $\propto \epsilon_\gamma^{-0.35} $ (see their Fig. 3). No spatial
profiles are given. \citet{NGC253:Domingo05} use for their main plot
$\tau_0 = 10$\,Myr and $\tau_\mathrm{ad}= 0.3$\,Myr, resulting in
$\tau_\mathrm{ad} < \tau_\mathrm{diff}$ for $\epsilon_\mathrm{p} \leq
1$\,TeV. Therefore, their SED should for low \g-ray energies be
advection-dominated and correspond to the source spectrum. The
corresponding curves in their Figs.~5 and 9 indicate this source
spectrum for an assumed particle source index $s \simeq 2.3$ up to
$\epsilon_\gamma$ equal to some hundreds of GeV. These curves indicate a
significant softening beyond about 1 TeV. For a smaller value of
$\tau_0$ their SED should fall off strongly $\propto
\epsilon_{\gamma}^{-(s-2)-0.5}$ already beyond some considerably lower value
of $\epsilon_\gamma$.

The above model results show hadronically dominated \g-ray emission up
to several TeV, where the \g\ rays are primarily produced in inelastic
collisions between nuclear CRs and target nuclei from the ambient ISM,
and subsequent $\pi^0$ decay. Bremsstrahlung and Inverse Compton
emission from primary and secondary electrons is not entirely
negligible and in some of these models only by a factor of a few below
the hadronic emission. This is in particular the case for \g-ray
energies below 100\,MeV, where the $\pi^0$-decay emission drops off,
and at the high-energy end, where the harder Inverse Compton emission
in the Thomson limit can win over the $\pi^0$-decay emission (if
electrons at these energies do not experience significant radiative
losses). Note however that the ratio of hadronic to leptonic emission
depends on the assumed source spectrum, the form of
$\tau_\mathrm{diff}(\epsilon_\mathrm{p})$, and on the assumed
electron-to-proton ratio \citep[which has to be $\gg$ than the
canonical value of 1/100 that electrons dominate over the hadronic
component and overcome energetics problems, see
e.g.][]{OhmHinton12}. Therefore it is very likely that the hadronic
emission dominates over the leptonic emission. That this is possible
is basically a consequence of the very high gas density $n_\mathrm{g}$
in the starburst region. \citet{HESS:NGC253_UL} find a $n_\mathrm{g}=
580$\,cm$^{-3}$ for a total gas mass of $6\times 10^7$\,$M_{\odot}$,
the value preferred by \citet{NGC253:Engelbracht98}. Lower and higher
gas masses as found by \citet{NGC253:Mauersberger96} and
\citet{NGC253:Sorai00}, respectively, imply an uncertainty of this
estimate of $\approx 20$\%. Note that this estimate only takes into
account the uncertainty in total gas mass and neglects any uncertainty
in the starburst volume due to possible projection effects. Clearly
this density is an average value -- the actual gas density is probably
extremely inhomogeneous, given the large localised energy inputs from
stars and in particular from their supernova explosions. Therefore,
the local density for p-p interactions could differ from the average
density. However, it is unlikely that protons escape from the denser
regions without significant losses. Adopting this $n_\mathrm{g}$
corresponds to an average loss time due to inelastic p-p collisions of
$\bar{\tau}_\mathrm{pp} \approx 1.1 \times 10^{5}$\,yr (this value is
mildly dependent on the observed spectral index due to the
energy-dependence of the p-p cross section $\sigma_\mathrm{pp}$; a
derivation of this value is given in the Appendix). Finally it has to
be noted that $\bar{\tau}_{\rm pp}$ scales inversely linear with the
inelasticity factor of the p-p collision (see Appendix). This can
introduce an additional uncertainty on $\bar{\tau}_{\rm pp}$ of
$\approx 20\%$.

We conclude that the \g-ray emission from the starburst region is
likely to be dominated by hadronic interactions. In the following we
will discuss the impact of system parameters such as SN rate, outflow
velocity and particle diffusion coefficient on the \g-ray emission
within the hadronic scenario. The possible additional role of discrete
\g-ray sources will be briefly discussed in
Section~\ref{sec:discrete}.

\subsection{Cosmic-ray escape}
\label{sec:energy_ind}

In the hadronically dominated case, there are three main scenarios for
the expected \g-ray emission, depending on the (potentially
energy-dependent) probability of cosmic rays escaping the starburst
region. In the case that the escape probability is $\ll 1$ at all
energies, the system can be said to be calorimetric. The measured
\g-ray flux is significantly lower than the flux expected in the
calorimetric limit for the canonical parameters, such as the
cosmic-ray acceleration efficiency as used in
Section~\ref{sec:cal}. Cosmic-ray escape must therefore be
considered. There are two competing mechanisms to remove cosmic rays
from the system: diffusion and advection. Dominance of advection
(which is an energy-independent process) over the full energy range
would result in a \g-ray spectrum that approximately resembles the
source spectrum. Diffusion on the other hand is an energy-dependent
process and would lead to a spectral steepening of the source
spectrum. At very low energies, the diffusion loss timescale
$\tau_\mathrm{diff}$ becomes very long and eventually comparable to
the advective loss timescale $\tau_\mathrm{ad}$. At the critical
energy, where the diffusive transport takes over from advective
transport and $\tau_\mathrm{diff} = \tau_\mathrm{ad}$, a spectral
break in the source spectrum is expected.

Empirically, for the Milky Way galaxy, the measurements of the CR
secondary-to-primary ratio \citep{Strong2007} suggest a softening of
the CR source spectrum by a factor $\propto E^{-\alpha}$, with $ 0.3
\lesssim \alpha \lesssim 0.6$, above a few GeV per nucleon as a result
of energy-dependent diffusion. This is consistent with a comparatively
modest production rate of hot gas and CRs per unit area of the
Galactic disk which allows most of the thermal gas to cool
radiatively. At the same time the excitation of magnetic field
fluctuations by the escaping CR particles and their coupling to the
thermal gas is equally modest. Although a CR-driven wind will develop
at large distances from the disk, the observed particle spectrum in
the Galactic disk is dominated by diffusive escape
\citep{Ptuskin97}. On the other hand, the wind from the starburst
region in \ngc\ is very strong as a result of very strong gas heating
that cannot be compensated by radiative cooling. The energy flux
density in CRs driving magnetic fluctuations is expected to be very
large as well, given the high SN rate in the small starburst volume
\citep[see e.g.][and Section~\ref{sec:bfield}]{HESS:NGC253_UL}. These
nonlinear effects massively diminish the role of particle diffusion
relative to global advection in the wind. Indeed, as shown in the next
Section, the observed \g-ray flux
$F_\mathrm{\gamma}^\mathrm{meas}(>200\,\mathrm{MeV})$ is well
explained quantitatively by pure particle advection in the wind from
the starburst nucleus, which requires a CR production $\Theta
E_\mathrm{SN} \simeq 10^{50}$\,erg per supernova
\citep[e.g.][]{Drury89} for the starburst parameters of \ngc.

The diffusion time in \ngc's starburst nucleus can be expressed as
$t_\mathrm{diff}=(H/2)^2 \kappa^{-1}$, where $\kappa$ denotes the
diffusion coefficient and $H \approx 60$\,pc is the height of the
starburst region, positioned symmetrically to the galactic
mid-plane. If $t_\mathrm{ad} \lesssim t_\mathrm{diff}$ is required,
then $\kappa$ should not exceed about $3\times
10^{27}$\,cm$^2$\,s$^{-1}$ for all energies below the last \hess\ flux
point at 4.7\,TeV that corresponds to a CR energy of $\approx 30$\,TeV
\citep[see e.g.][]{Kelner2006}. Such a small diffusion coefficient
could be the result of strong wave excitation by the exceedingly
concentrated CR production in the small starburst volume
\citep{HESS:NGC253_UL}. This value can be compared to that of Bohm
diffusion which should be considered the slowest possible form of
diffusion for a randomised magnetic field configuration. The Bohm
limit is given by $\kappa_\mathrm{Bohm} \approx 3\times
10^{22}\,(\epsilon_\mathrm{GeV}/B_\mathrm{\mu G})$\,cm$^2$\,s$^{-1}$,
where $\epsilon_\mathrm{GeV}$ is the CR energy in GeV and
$B_\mathrm{\mu G}$ is the magnetic field in $\mu$G. For a particle
energy of 30\,TeV and a magnetic field of 100\,$\mu$G, as estimated in
the next section, the Bohm diffusion coefficient is about $3\times
10^{24}$\,cm$^2$\,s$^{-1}$ and thus still three orders of magnitude
smaller than the maximum diffusion coefficient deduced from the single
power-law spectrum of the \g-ray data. The upper limit to the
diffusion coefficient estimated for the starburst region in \ngc\ can
also be compared to the average value inferred for the Milky Way
galaxy which is given by $\kappa_\mathrm{gal} \approx 1.5 \times
10^{30} (E/1\,\mathrm{TeV})^{1.1}$\,cm$^2$\,s$^{-1}$
\citep{Ptuskin97,HESS:NGC253_UL}. With such a comparison,
extrapolating to a particle energy of 30\,TeV, $\kappa \lesssim 5
\times 10^{-5} \kappa_\mathrm{gal}$ is found. This is an interesting
requirement on the scattering strength of the magnetic field
fluctuations\footnote{This value can also be compared to the
  gyroradius of a proton with 1\,TeV energy in a $100\,\mu$G field,
  which is $r_g \approx 3.3 \times 10^{13} (\epsilon_{\mathrm{TeV}} /
  B_{\mathrm{100\,\mu G}})\,\mathrm{cm} \approx 2\,\mathrm{A.U.}$} in
the starburst region, but certainly not an outrageous one, comparing
with the Bohm scattering level.

Dominant transport by advection is suggested by the observation of a
rather hard \g-ray spectrum and the smooth alignment of the spectrum
in the HE and the VHE \g-ray regime --- the most noticeable
observational result of this paper (see
Section~\ref{sec:jointfit}). Therefore, taking the combined {\it
  Fermi} and \hess\ spectra as single power law as indicated in
Fig.~\ref{fig:ngc_spectrum} with differential index
$\Gamma_\mathrm{sim} = 2.34$, argues that energy-dependent diffusion
is not important in the energy range covered
\footnote{Energy-independent diffusion can in principle occur as a
  result of large-scale turbulent motions in the gas and may also
  contribute to energetic-particle confinement in the gas that is
  systematically streaming with $v_\mathrm{wind}$
  \citep[e.g.][]{Bykov01,Parizot04}. This effect is not included here
  explicitly.}. In fact, advection alone already explains the
magnitude of $F_{\gamma}^\mathrm{meas}$ quantitatively.
  
Therefore, two conclusions may be drawn here: (i) the observed
spectrum is most likely the result of an energy-independent transport
mechanism, i.e. advection and adiabatic expansion in the starburst
wind and inelastic nuclear energy losses, and (ii) the combined
observed \g-ray spectrum in the HE and VHE \g-ray regimes might in
this case correspond approximately to the mean spectrum of the
ensemble of CR sources in the starburst region \footnote{In fact, the
  \g-ray spectrum is slightly harder than the parent proton spectrum
  \citep{Kelner2006,Kamae2006,Karlsson2008}.}.

\subsection{\g-ray flux estimate}

If, as argued in the previous section, CR escape from the starburst
region is independent of particle energy, the expected total \g-ray
energy flux $F_{\gamma}^{\mathrm{exp}}$ can be estimated simply from
the parameters of the system. This independence is approximately true
also for the loss rate with respect to inelastic p-p collisions; in
fact, this loss rate will be approximated by an average over the range
of particle energies corresponding to the observed $\gamma$-ray
spectrum (see Appendix). In this case the particle transport equation
can be simply integrated over the energy range of the pion-producing
particles. In a leaky box-type approximation for the starburst region
\citep[e.g.][]{Berezinskii90} the result can in addition be integrated
over the starburst volume. Given that the lifetime of the starburst is
about $(2 - 3) \times 10^7$\,yr \citep[see][]{NGC253:Engelbracht98}
which is large compared to the advective loss time of about
$10^5$\,yrs, the system is in a quasi-steady state. This results in
the following balance relation for the total non-thermal energy
$E_\mathrm{pp}^{\pi}$\, of pion-producing particles in the starburst
region:

\begin{equation} 
  E_\mathrm{pp}^{\pi} \left[{1 \over \tau_\mathrm{ad}} + {1 \over
      \tau_\mathrm{adiab}} + {1 \over \bar{\tau}_\mathrm{pp}} \right] \approx Q^{\pi},
\label{eq1}
\end{equation}
\noindent In the derivation of Eq.~\ref{eq1} the source spectrum and
the resulting spectrum of pion-producing particles in the starburst
region are both assumed to have approximately the form of a power-law
spectrum in momentum $\propto p^{-(\Gamma_\mathrm{sim}+2)}$, as
expected from the theory of diffusive shock acceleration
\citep[e.g.][see Appendix]{Blandford87}.

In Eq.~\ref{eq1} the quantity $Q^{\pi}= f_{\pi} Q$ is the fraction
$f_{\pi} <1$, due to pion-producing particles, of the total input rate
Q of non-thermal energy from the CR sources. Approximately $f_{\pi}
\approx 3-\Gamma_\mathrm{sim} = 0.66$, assuming that the observed
$\gamma$-ray spectrum with $\Gamma_\mathrm{sim} = 2.34$ resembles the
source spectrum. The quantity $\tau_\mathrm{ad} =
(H/2)/v_\mathrm{wind} \approx 10^{5}$\,yr denotes the advective loss
time, where $v_\mathrm{wind} \approx 300$\,km\,s$^{-1}$
\citep{NGC253:Zirakashvili06} is the velocity of the starburst wind at
the top/bottom of the starburst region. The adiabatic loss time is
given by $\tau_\mathrm{adiab}$ in the accelerating outflow. In a first
approximation the flow speed $\vec{V}$ rises from zero at the galactic
symmetry plane in the perpendicular direction $z$ as $|\vec{V}| =
v_\mathrm{wind} [|z|/(H/2)]$\,km\,s$^{-1}$, yielding an adiabatic loss
rate $1/\tau_\mathrm{adiab} \approx
(\Gamma_\mathrm{sim}-1)\nabla{\vec{V}}/3 = (\Gamma_\mathrm{sim}-1)/(3
\tau_\mathrm{ad})$. Finally, $\bar{\tau}_\mathrm{pp} \simeq
(n_\mathrm{g} \, 0.5 \, c \, \langle \sigma_\mathrm{pp}
\rangle)^{-1}$, is an average energy loss time for inelastic,
catastrophic proton-proton collisions, which is inversely proportional
to the effective gas density $n_\mathrm{g}$. The inelasticity is taken
as 0.5. With the mean cross section $\langle\sigma_\mathrm{pp}\rangle
\, \approx 33$\,mb (see Appendix) this leads to a mean collisional
energy loss time $\bar{\tau}_\mathrm{pp} \simeq 1.1 \times
10^5\,(n_\mathrm{g}/580\,\mathrm{cm}^{-3})^{-1}$\,yr.

The total hadronic $\gamma$-ray energy flux density
$F_{\gamma}^{\mathrm{exp}}$ and $E_\mathrm{pp}^{\pi}$ are connected by

  \begin{equation}
    F_{\gamma}^{\mathrm{exp}} \approx {E_\mathrm{pp}^{\pi} \eta \over 4 \pi d^2 \bar{\tau}_\mathrm{pp}},
\label{eq2}
\end{equation} 
where $\eta \simeq 1/3$ is the $\pi^0$-fraction from overall
pion-production through hadronic collisions.

Eq.~\ref{eq2} assumes an optically thin \g-ray emission region. Indeed
the optical depth $\tau_\mathrm{\gamma\gamma}$ for
$\gamma\gamma$-absorption in the diffuse radiation field in the
starburst region and the remaining part of \ngc\ is small compared to
unity for the $\gamma$-ray energies considered,
i.e. $\tau_{\gamma\gamma} < 0.1$ for $\epsilon_\gamma < 2$\,TeV
\citep[see][]{Inoue11}.

Assuming that the CR energy sources in the starburst region are the SN
remnants, then $Q = \nu_\mathrm{SN} \Theta E_\mathrm{SN}$, where
$\nu_\mathrm{SN}$ is the SN rate, $\Theta < 1$ is the CR production
efficiency, and $E_\mathrm{SN}$ is the total hydrodynamic energy
release per event. In order to comply with the overall energetics of
CRs in the Milky Way galaxy, an average energy release $\Theta
E_\mathrm{SN} \approx 10^{50}$\,erg into nuclear CRs should be assumed
\citep{Drury89}, with an uncertainty of about a factor of 2. For
$E_\mathrm{SN}= 10^{51}$\,erg this implies $\Theta \approx 0.1$.

It is of interest to compare this expected flux with the measured flux
$F_\mathrm{\gamma}^\mathrm{meas}(> 200\mathrm{MeV})\approx 5.3 \times
10^{-12}$\,erg cm$^{-2}$ s$^{-1}$. For this purpose a nominal value of
$0.03$\,yr$^{-1}$ for $ \nu_\mathrm{SN}$ is assumed
\citep{NGC253:Engelbracht98}. Inserting $E_\mathrm{pp}^{\pi}$ from
Eq.~\ref{eq1} into Eq.~\ref{eq2} leads to\footnote{The assumption that
  the parent proton spectrum and the resulting \g-ray spectrum have
  the same power-law index infers an error of $\approx 20$\% in
  $E_{pp}^{\pi}$ (see also Appendix).}

\begin{eqnarray}
  \label{eq3}
  % roughly-equals
  F_\mathrm{\gamma}^{\mathrm{exp}} & \simeq & 2.6 \times 10^{-11} \, 
  \mathrm{erg}\,\mathrm{cm}^{-2}\,\mathrm{s}^{-1} \nonumber\\
  & \times & \left[\left({E_\mathrm{SN} \over {10^{51}\mathrm{erg}}}\right) 
    \left({\Theta \over {0.1}}\right)
%    \left({\eta \over 1/3}\right)
    \left({\nu_\mathrm{SN} \over {0.03\mathrm{yr}^{-1}}}\right) 
    \left({f_{\pi} \over 0.66}\right)
    \left({d \over {2.6\mathrm{Mpc}}}\right)^{-2}\right]\nonumber\\
  & \times &
  \left[{3 \over {(\bar{\tau}_\mathrm{pp}/1.1\times 10^5\,\mathrm{yr})
        / (\tau_\mathrm{ad}/{10^5\,\mathrm{yr}}) \times 1.1 \Gamma_\mathrm{sim} +
        3}}\right],
\end{eqnarray}

\noindent where all parameters like
$(\tau_\mathrm{ad}/{10^5\,\mathrm{yr}})$ are written in terms of
their nominal values.  In the following the first bracket on the
r. h. s. of Eq.~\ref{eq3} is referred to as the {\it production term}
and the second bracket is referred to as the {\it loss term} of
Eq.~\ref{eq3}.

Considering nominal parameter values in Eq.~\ref{eq3}, the expected
flux $F_\mathrm{\gamma}^{\mathrm{exp}} \simeq 10^{-11} \,
\mathrm{erg}\,\mathrm{cm}^{-2}\,\mathrm{s}^{-1}$, and then
$F_\mathrm{\gamma}^{\mathrm{exp}} /F_\mathrm{\gamma}^\mathrm{meas} \approx 1.9$. Given
the uncertainties in the measurements of the numerous multi-wavelength
parameters involved the two fluxes are quite close.  This supports the
general picture on which Eq.~\ref{eq3} is based.

The question is, to which extent this result leads to physically
relevant bounds on physics quantities like the efficiency parameter
$\Theta E_\mathrm{SN}$. There are two possibilities to reduce the
expected $F_{\gamma}^{\mathrm{exp}}$ to the observed value
$F_\mathrm{\gamma}^\mathrm{meas}$. The first of them can be achieved
by reducing $\Theta E_\mathrm{SN}$ compared to its value inferred for
the Milky Way galaxy and/or by reducing the astrophysical parameter
$\nu_\mathrm{SN} /d^2$, i.e. by reducing the CR {\it production term}
in Eq.~\ref{eq3}. The second possibility is to increase the ratio
$\bar{\tau}_\mathrm{pp} / \tau_\mathrm{ad}$ either by decreasing the
effective gas density $n_\mathrm{g}$ seen by the pion-producing
particles during their advective escape, relative to the average gas
density observed, or by increasing $v_\mathrm{wind}$. In this second
case the source term could be kept at its nominal value, especially
$\Theta E_\mathrm{SN}$ could be kept at the value of
$10^{50}$\,erg. Decreasing $\bar{\tau}_\mathrm{pp} /
\tau_\mathrm{ad}$ relative to its nominal value, on the other hand,
is quite implausible for two reasons: firstly, the wind velocity
cannot become significantly smaller without causing difficulties to
explain the spatial extent of the starburst region as seen in radio
observations \citep{NGC253:Zirakashvili06}. Secondly, the total gas
density cannot significantly increase, given the observations of the
molecular gas mentioned earlier. This means that the {\it loss term}
in Eq.~\ref{eq3} has a maximum value for the default parameters in
Eq.~\ref{eq3}. As a consequence the product $\nu_\mathrm{SN} \Theta
E_\mathrm{SN}$ should at most decrease by a factor 1/1.9, if at all,
from its value $10^{50}\,\mathrm{erg} \times 0.03\, \mathrm{yr}^{-1}$
for the nominal parameters. Effectively, $\Theta E_\mathrm{SN}$
depends on the value $\nu_\mathrm{SN}$. Only for $\nu_\mathrm{SN} \gg
0.03\, \mathrm{yr}^{-1}$ should $\Theta E_\mathrm{SN}$ on average be
small compared to $10^{50}\,\mathrm{erg}$, i.e. in contrast to the
average situation in the Milky Way galaxy and theoretical expectations
for individual supernova remnants there.

To derive the quantities of \ngc\ the reference distance of $d =
2.6$\,Mpc has been used most widely in the literature. However, as
discussed in the Introduction, this distance has recently been revised
to $d = 3.5$\,Mpc.  Therefore, an appropriate scaling of the
astronomical parameters in Eq.~\ref{eq3} needs to be
considered. Indeed, the total gas mass as determined from the CO line
flux scales as $d^2$, whereas the starburst volume scales as
$d^3$. This implies that the gas density scales as $d^{-1}$ and
$\bar{\tau}_\mathrm{pp} \propto d$. The supernova rate is derived from
the FIR continuum flux and scales as $\nu_\mathrm{SN} \propto d^{2}$.
The wind velocity is derived to be consistent with the geometry of the
radio brightness distribution yielding $v_\mathrm{wind} \propto d$. As
a consequence $\tau_\mathrm{ad}$ is independent of $d$. For
$d=3.5$\,Mpc the nominal value of $F_\mathrm{\gamma}^{\mathrm{exp}}$
in Eq.~\ref{eq3} is therefore about $8.3\times
10^{-12}$\,erg\,cm$^{-2}$\,s$^{-1}$.  Following the above physical
arguments and changing the distance to $d=3.5$\,Mpc, the production
term on the r.h.s. of Eq.~\ref{eq3} must not become smaller by more
than a nominal factor $\simeq 1/1.6$, in order to reduce
$F_\mathrm{\gamma}^{\mathrm{exp}}$ to
$F_\mathrm{\gamma}^{\mathrm{meas}}$.

In general, the simple model presented here agrees quite well with the
observed values, given the observational uncertainties of the
astronomical parameters and the possibility that for SN remnants in
such a dense medium the total non-thermal energy generated per event
$\Theta E_\mathrm{SN}$ might indeed be lower by a factor $\simeq
1/1.6$ than typically assumed for an object in the average ISM of the
Milky Way galaxy.

\subsection{Hadronic calorimetry}
\label{sec:cal}
In this scenario, to find the extent to which the starburst region
behaves calorimetrically in the presence of advective and diffusive
escape, the total energy production rate in hadronic collisions
$L_\mathrm{coll} = 4 \pi d^2 F_{\gamma}^\mathrm{meas} / \eta $ is
compared with the total production $L_\mathrm{CR}(\pi)=
f_{\pi}\nu_\mathrm{SN} \Theta E_\mathrm{SN}$ of CRs capable of
producing hadronic \g\ rays. Such a comparison for \ngc\ has
previously been made by \citet{HESS:NGC253_UL,Loeb06}, and
\citet{Thompson07}. $L_\mathrm{CR}(\pi)$ depends on the fraction of
energy available for pion-production $f_\pi \approx 3 - s$, with $s$
being the source spectral index. In the present scenario $f_\pi \simeq
\Gamma_\mathrm{sim}$ and therefore $f_\pi = 0.66$.

Adopting a value of 0.03\,yr$^{-1}$ for $\nu_\mathrm{SN}$
\citep{NGC253:Engelbracht98} and assuming for $\Theta E_\mathrm{SN}
\approx 10^{50}$\,erg again results in

\begin{eqnarray}
  \frac{L_\mathrm{coll}}{L_\mathrm{CR}(\pi)}  \approx  0.21 
  \times  \left(\frac{F_{\gamma}^{\mathrm{meas}}} {5.3 \times 10^{-12}
      \,\mathrm{erg}\,\mathrm{cm}^{-2}\,\mathrm{s}^{-1}}\right) \times \nonumber\\
  \left[\left({E_\mathrm{SN} \over {10^{51}\mathrm{erg}}}\right) 
    \left({\Theta \over {0.1}}\right)
%    \left({\eta \over 1/3}\right)
    \left({\nu_\mathrm{SN} \over {0.03\mathrm{yr}^{-1}}}\right) 
    \left({f_{\pi} \over 0.66}\right)
    \left({d \over {2.6\mathrm{Mpc}}}\right)^{-2}\right]^{-1}\,.
\label{eq4}
\end{eqnarray}

\noindent This fraction of about 20 percent is then a measure of the
extent to which the starburst region is calorimetric with respect to
its hadronic interactions, dissipating its own non-thermal output; and
since $\nu_{\mathrm{SN}}$ is proportional to the FIR luminosity and
hence to $d^2$, this result is formally
independent of the distance $d$. The ratio
$L_\mathrm{coll}/L_\mathrm{CR}(\pi)$ is comparable to the value found
by \citet{NGC253:Lacki11}. Like $F_\mathrm{\gamma}^{\mathrm{exp}} /
F_\mathrm{\gamma}^{\mathrm{meas}}$, it is affected by the
uncertainties of the input parameters. Adopting a larger ratio of
$\nu_\mathrm{SN} /d^2$ would decrease this value, whereas a lower
efficiency $\Theta E_\mathrm{SN}$ would increase it. The difference
between $F_\mathrm{\gamma}^{\mathrm{exp}}$ and
$F_\mathrm{\gamma}^{\mathrm{meas}}$, discussed above, shows that the
value of the production term of Eq.~\ref{eq3} should possibly be
smaller than unity but not smaller than $\simeq 1/1.6$ for $d =
3.5$\,Mpc. This suggests that the calorimetric fraction may be larger
than 20 percent, possibly reaching up to about 30 percent.

\subsection{Cosmic-ray energy density and magnetic field strength}
\label{sec:bfield}

Within this framework the non-thermal energy density
$U_\mathrm{pp}^{\pi}$ of the $\pi^0$-producing particles in the
starburst region can be simply calculated from Eq.~\ref{eq2} by
substituting $F_{\gamma}^{\mathrm{exp}} = F_\gamma^\mathrm{meas}$ and
dividing the resulting $E_\mathrm{pp}^{\pi}$ by the estimated
volume. This gives $U_\mathrm{pp} \approx 230$\,eV\,cm$^{-3}$,
independent of $d$. This is a lower limit for the total non-thermal
energy density $U_\mathrm{pp}$. Since it is difficult to estimate the
contribution of the lower-energy particles produced in the sources,
because of their poorly determined ionisation losses in the outflow,
one can only give an upper limit $U_\mathrm{pp} < U_\mathrm{pp}^{\pi}
/f_{\pi} \approx 340$\,eV\,cm$^{-3}$. The values in this range are
more than a hundred times larger than the CR energy density in the
Milky Way galaxy and, by implication, on average in the disk of \ngc
\, (see Introduction). Assuming, as for the Milky Way, a CR scale
length of about 1 kpc for the extended disk of \ngc\ in comparison to
the $\approx 40$\,pc gradient scale in the starburst region (for $d =
3.5$\,Mpc), the latter's CR pressure gradient is almost 4 orders of
magnitude larger than that on average in the disk.

Equating $U_\mathrm{pp}$ with the magnetic field energy density
$\vec{B}^2 / 8\pi$ in the sense of a conventional equipartition
argument for magnetic field and CRs, results in a range for the
r.m.s. magnetic field strength $90\,\mu\mathrm{G} \lesssim
B_\mathrm{eq} \lesssim 120\,\mu$G. Note that the estimates of
$U_\mathrm{pp}$ and $B_\mathrm{eq}$ are also affected by the
uncertainty of the inelasticity factor for p-p collisions that could
be $\approx 20\%$ lower \citep{PPinelas03}. This is to be compared
with the $(160 \pm 20)\,\mu$G equipartition field strength for the
so-called nuclear region, estimated recently by \citet{Heesen11} from
radio continuum observations and assuming a proton to electron ratio
of 100.

\subsection{Contribution of discrete sources}
\label{sec:discrete}

It has recently been suggested that the HE \g-ray emission from \ngc\
might indeed come from interactions of diffuse nuclear CRs with the
ambient gas, but that this contribution should substantially diminish
at energies $\geq 10$\,GeV as a result of energy-dependent diffusion
losses from the starburst region \citep{NGC253:Mannheim12}. The
observed TeV emission was argued to be due to Inverse Compton emission
from unresolved pulsar wind nebulae (PWN) instead, i.e. from a
separate population of discrete sources. For such PWNe the evolution
was assumed to be the same in the disk of the Milky Way galaxy and in
the extremely high-density environment of the starburst nucleus of
\ngc. It has however to be noted that electron cooling times in
typical starburst environments are very short and electrons are
expected to cool on timescales of a few hundred years
\citep{OhmHinton12}.

While in the present paper it is argued that advective removal should
dominate over diffusive losses of energetic nuclear particles deep
into the TeV range, a discussion of the role of PWNe in such an
environment is beyond the scope of the present work.  On the other
hand, the observations show that the spectrum of \ngc\ can be well
described by a single power law over the combined HE and VHE energy
range, disfavouring two distinct spectral components and suggesting
that the same physical processes dominate over the entire energy range
in question. Therefore, additional discrete \g-ray sources seem to
play a minor role.

%%%%%%%%%%%%%%%%%%%%%%%%%%%%%%%%%%%%%%%%%%%%%%%%%%%%%%%%%
%
%    Section V : Conclusions 
%
%%%%%%%%%%%%%%%%%%%%%%%%%%%%%%%%%%%%%%%%%%%%%%%%%%%%%%%%%
\section{Summary and outlook}

%%% HE and VHE results
The analysis results of 177 hours of \hess\ data obtained in
observations of the starburst galaxy \ngc \, are reported. The
reconstructed energy spectrum is best described by a po wer law with
differential photon index $\Gamma=2.14 \pm 0.18_{\mathrm{stat}} \pm
0.30_{\mathrm{sys}}$ and differential flux normalisation at 1\,TeV of
$F_0$ = $(9.6 \pm 1.5_{\mrm{stat}} ~(+5.7,-2.9)_\mrm{sys}) \times
10^{-14}\UNITS{TeV^{-1}\,cm^{-2}\,s^{-1}}$. In addition to the \hess\
data, the analysis of the 30 months {\it Fermi}-LAT data set revealed
an improved best fit position compatible with the optical centre of
the galaxy and with the \hess\ source within statistical errors. The
reconstructed differential photon index is
$\Gamma=2.24\pm0.14_{\mathrm{stat}}\pm0.03_{\mathrm{sys}}$ and the
integral flux between $(0.2-200)$\,GeV is $F(0.2 - 200\UNITS{GeV}) =
(4.9\pm1.0_{\mrm{stat}}\pm0.3_{\mrm{sys}})\times
10^{-9}\UNITS{cm^{-2}\,s^{-1}}$. The HE and VHE \g-ray spectra of the
starburst region can be described by a simultaneous power-law fit with
differential photon index $\Gamma_\mathrm{sim}=2.34\pm0.03$ and a fit
probability of 30\%. This result implies that no spectral break or
turnover is required to explain the \g-ray data. The corresponding
total energy flux density corresponds to $F^E(>200\,\mathrm{MeV})
\approx 5.3 \times 10^{-12}$\,erg\,cm$^{-2}$\,s$^{-1}$.  Assuming the
remaining disk of \ngc \, to be quantitatively similar to the Milky
Way galaxy, the starburst region outshines the rest of \ngc\ by an
order of magnitude in HE and VHE $\gamma$ rays, consistent with the
detection of the object as a H.E.S.S. point source.

%%% Discussion and implications for advective and diffusive losses
Model predictions which assume a dominantly hadronic origin of the
\g-ray emission are roughly compatible with the spectral results
presented in this work. For a set of reasonable parameters the CR
energy, which is lost in p-p interactions and partly re-appears in
$\pi^0$-decay \g-ray production, is inferred in the present work as
$\approx 20$\% and possibly up to $\approx 30$\% of the total
non-thermal energy produced in the starburst region, assuming a
distance of $3.5$\,Mpc and a 10\% efficiency for cosmic-ray
acceleration in starburst supernova remnants. Note however, that the
multi-wavelength observables are only known within a considerable
error margin, which can change these percentages significantly. CRs
are also removed by diffusion and advection from the starburst
region. Since the former process is energy dependent, a spectral
steepening with energy would be expected. The smooth alignment of the
HE and VHE \g-ray spectra over four decades in energy hence indicates
that advective losses in \ngc\ most likely dominate from a few GeV to
more than 10\,TeV. Even at such high energies the diffusion
coefficient would still be more than two orders of magnitude larger
than the Bohm diffusion coefficient. It therefore seems likely that
the observed spectrum can be characterised by the same photon index as
the average particle accelerator in the starburst region.

The form of the \g-ray spectrum of NGC 253 can be compared with
another starburst galaxy, M\,82 in the Northern Hemisphere, detected
by the {\it Fermi}-LAT \citep{Fermi:NGC253M82} and VERITAS
\citep{VERITAS:M82} collaborations, respectively. If one looks at the
corresponding HE and VHE \g-ray spectra of M\,82, the overall shape
looks rather similar to the one presented in this paper for
NGC\,253. Even though the starburst in M\,82 is in all probability
triggered by the interaction with the companion galaxy M\,81, the
spectral similarity is consistent with the assumption that the CR
sources in both galaxies produce similar energetic particle spectra
and may therefore be of the same nature.

\begin{acknowledgements}
  We thank S.R. Kelner for valuable discussions and the anonymous
  referee for his very useful comments. The support of the Namibian
  authorities and of the University of Namibia in facilitating the
  construction and operation of H.E.S.S. is gratefully acknowledged,
  as is the support by the German Ministry for Education and Research
  (BMBF), the Max Planck Society, the German Research Foundation
  (DFG), the French Ministry for Research, the CNRS-IN2P3 and the
  Astroparticle Interdisciplinary Programme of the CNRS, the
  U.K. Science and Technology Facilities Council (STFC), the IPNP of
  the Charles University, the Czech Science Foundation, the Polish
  Ministry of Science and Higher Education, the South African
  Department of Science and Technology and National Research
  Foundation, and by the University of Namibia. We appreciate the
  excellent work of the technical support staff in Berlin, Durham,
  Hamburg, Heidelberg, Palaiseau, Paris, Saclay, and in Namibia in the
  construction and operation of the equipment. SO acknowledges the
  support of the Humboldt foundation by a Feodor-Lynen research
  fellowship.
\end{acknowledgements}

\appendix
\section{}
Eq.~\ref{eq1} of the main text can be derived from the transport
equation for the isotropic part $f(\vec{x},p,t)$ of the particle
momentum distribution in a simple model, neglecting the speed of the
scattering fluctuations compared to the fluid mass velocity $\vec{V}$
as well as diffusion in momentum space \citep[e.g.][]{Blandford87}:

\begin{eqnarray}
  \tilde{Q}(\vec{x},p,t)  =  {\partial f\over \partial t} - \nabla (\kappa \nabla
  f) + \nabla (\vec{V} f) -\frac{1}{p^2}\frac{\partial}{\partial p}
  \left(p^2 \frac{p}{3}(\nabla \vec{V}) f \right)
  -  {f\over {\bar{\tau}_\mathrm{pp}}},
\label{eqA1}
\end{eqnarray}

\noindent where $\kappa(p)$ denotes the spatial diffusion coefficient,
$\bar{\tau}_\mathrm{pp}$ is the average energy loss time as a result
of inelastic, catastrophic nuclear collisions (assumed to be
energy-independent for $p > 2m_\mathrm{p}c$ and infinite for $p <
2m_\mathrm{p}c$, see below), and $\tilde{Q}(\vec{x},p,t)$ is the
particle production rate of the CR sources.  The adiabatic loss rate
of a particle in the accelerating flow is given by $ p/3 \nabla
\vec{V}$. In the energy range of pion-producing nuclei ionisation
losses are neglected. Assuming a steady state ($\partial f / \partial
t = 0$) and neglecting particle diffusion ($\kappa = 0$), the flow
velocity $\vec{V}$ is approximated to be perpendicular to the galactic
disk mid-plane, varying as $\vec{V} = (0,0, v_\mathrm{wind} \times z
(H/2)^{-1})$ for $0<|z|<H/2$, where the constant parameter
$v_\mathrm{wind}$ is the wind velocity at the starburst boundary.

To obtain the balance relation for the kinetic energy of the
pion-producing CRs $E_\mathrm{pp}^{\pi}= \int_{V_\mathrm{SB}}d^3x \,
4\pi \int_{p_\mathrm{min}}^{p_\mathrm{max}} dp p^2 E_\mathrm{kin} f $
in the starburst region, Eq.~\ref{eqA1} is multiplied by the particle
kinetic energy $E_\mathrm{kin}$ and integrated over the relevant
momenta as well as over the spatial starburst volume
$V_\mathrm{SB}$. In a leaky box approximation for the starburst region
\citep[e.g.][]{Berezinskii90}, $f(\vec{x},p)$ and
$\bar{\tau}_\mathrm{pp}$ are assumed to be spatially uniform. It is
then clear for this particle distribution that it has the same
momentum dependence as the spatially averaged source production rate
$\int_{V_\mathrm{SB}}d^3x \tilde{Q}(\vec{x},p)/V_\mathrm{SB}$ for
$p_\mathrm{min}< p < p_\mathrm{max}$. In addition, it is assumed that
the production spectrum in the sources is not strongly influenced by
energy-dependent losses in the sources themselves. The averaged
particle source production rate can then be assumed to have a
power-law dependence $\propto p^{-(s + 2)}$ in particle momentum $p$
for all momenta above the injection energy, consistent with an origin
of the CR particles from diffusive shock acceleration
\citep[e.g.][]{Blandford87}. And $s \approx \Gamma_\mathrm{sim}$,
where $\Gamma_\mathrm{sim}= 2.34$ is the power-law index of the
observed $\gamma$-ray flux. The minimum momentum $p_\mathrm{min}$ in
Eq.~\ref{eqA1} roughly corresponds to GeV energies. Here
$p_\mathrm{min}$ is taken as $2m_\mathrm{p}c$. Since the maximum
observed $\gamma$-ray energy is 4.7 TeV, corresponding to a proton
energy of $cp_\mathrm{max} \approx 30$\,TeV, momenta $p>
p_\mathrm{max}$ make a negligible contribution. Therefore, it is
possible to put $p_\mathrm{max} = \infty$.

In order to obtain an analytical estimate the above integral for
$E_\mathrm{pp}^{\pi}$ is approximated by using the relativistic
formula $E_\mathrm{kin} = cp$, for $p > 2m_\mathrm{p}c$
\citep{Drury89}.

Integrating the individual terms in Eq.~\ref{eqA1} by parts over momentum
results in :

\begin{equation}
  E_\mathrm{pp}^{\pi} \left[\frac{1}{\tau_\mathrm{ad}} +
    \frac{1}{\tau_\mathrm{adiab}} + \frac{1}{\bar{\tau}_\mathrm{pp}}\right] \approx
  Q^{\pi},
\label{eqA2}
\end{equation}

\noindent where $\tau_\mathrm{ad} = (H/2)/v_\mathrm{wind}$,
$\tau_\mathrm{adiab}\approx 3 \tau_\mathrm{ad}/(\Gamma_\mathrm{sim} -
1)$, and $Q^{\pi} \approx 4\pi \int_{2m_\mathrm{p}c}^{\infty} dp p^2
\tilde{Q}$. This quantity $ Q^{\pi} = f_{\pi} Q$ defines the fraction
$f_{\pi} < 1$ of the total particle energy production $Q =
\int_{V_\mathrm{SB}}d^3x \, 4\pi \int_{0}^{\infty} dp p^2 E_\mathrm{kin}
\tilde{Q}$, integrated over all momenta, that is available for pion
production. Consistent with the approximation for $Q^{\pi}$ the
integral for $Q$ is approximated by using the newtonian formula
$E_\mathrm{kin}=p^2/2m_\mathrm{p}$ for $p < 2m_\mathrm{p}c$ and the
relativistic formula $E_\mathrm{kin} = cp$ for $p > 2m_\mathrm{p}c$
\citep{Drury89}.

Assuming $2 \lesssim s \lesssim 3$, $p_\mathrm{inj}
\ll 2m_\mathrm{p}c$, for the injection momentum $p_\mathrm{inj}$, and
$p_\mathrm{max} \gg 2m_\mathrm{p}c$, then gives:

\begin{equation}
  Q^{\pi} \approx 4\pi Q_0 \int_{2m_\mathrm{p}c}^{p_\mathrm{max}} dp p^2 cp
  (p / 2m_\mathrm{p}c)^{-s - 2} \approx {4\pi Q_{0}c /
    (s -2)}(2m_\mathrm{p}c)^4
\label{eqA3}
\end{equation}

\noindent and

\begin{equation}
  Q - Q^{\pi} \approx  4\pi Q_0 \int_{p_\mathrm{min}}^{2m_\mathrm{p}c} dp p^2
  {p^2 / 2m_\mathrm{p}} (p / 2m_\mathrm{p}c)^{- s - 2}
  \approx {4\pi Q_{0}c / (3-s)}(2m_\mathrm{p}c)^4 .
\label{eqA4}
\end{equation}

Therefore, $Q^{\pi} \approx (3-s)Q$ and $f_{\pi} \approx 3-s$ for the assumed
momentum spectrum. For $s = \Gamma_\mathrm{sim}= 2.34$, $f_{\pi} \approx
0.66$. For a very hard source spectrum with $s$ close to 2, $f_{\pi}$ would be
close to 1. Eq.~\ref{eqA2} corresponds to Eq.~\ref{eq1} of the main text.

The approximate quantity $\bar{\tau}_\mathrm{pp}$\,, i.e. the mean
loss time that appears in Eq.~\ref{eqA1}, arises from averaging the
energy loss rate $\tau_\mathrm{pp}^{-1}(E)$ over the differential
energy spectrum of the colliding particles. Approximating the
differential energy dependence of the particle spectrum by that of the
observed $\gamma$-ray spectrum implies that the average can be taken
over a power law spectrum $\tilde{f}(E_\mathrm{kin}) \propto
E_\mathrm{kin}^{-\Gamma_\mathrm{sim}}$ which has the same index
$\Gamma_\mathrm{sim} = 2.34$ as the differential $\gamma$-ray flux as
a function of the $\gamma$-ray energy $\epsilon_{\gamma}$ in the energy
region observed for \ngc, thus defining:

\begin{equation}
  \frac{1}{\bar{\tau}_\mathrm{pp}} =
  \frac{\int dE E  \tilde{f}(E)/ \tau_\mathrm{pp}}
  {\int dE E \tilde{f}(E)}
\label{eqA5}
\end{equation}

\noindent The integration limits are the particle (proton) energies
corresponding to the observed energy range of the $\gamma$-ray energy
spectrum. Since $\tau_{pp}^{-1} = n_\mathrm{g} c f_\mathrm{in}
\sigma_\mathrm{pp}(E_\mathrm{kin})$ with the total cross section
$\sigma_\mathrm{pp}$ for inelastic pp-collisions and inelasticity
factor $f_\mathrm{in} \approx 0.5$ , cf. \citet{Book:Aharonian04},
this amounts to the following average over the inelastic nuclear
cross-section $\sigma_\mathrm{pp}$:

\begin{equation}
  \frac{1}{\bar{\tau}_{pp}} = \frac{n_\mathrm{g} c f_\mathrm{in}
    \int dE\,E\,\sigma_\mathrm{pp}(E)\, \tilde{f}(E)}
  {\int dE \, E\,\tilde{f}(E)}.
\label{eqA6}
\end{equation}

\noindent It is a convenient approximation to choose the integration limits as
$E_\mathrm{kin}^{\mathrm{min}} = 1$\,GeV, and as
$E_\mathrm{kin}^{\mathrm{max}}\approx (4.7\, \mathrm{TeV}/0.17) \approx 30$\,TeV
according to the maximum $\gamma$-ray energy $\epsilon_{\gamma}= 4.7$\,TeV
observed. For $E_\mathrm{kin} > 1$\,GeV the cross section $\sigma_\mathrm{pp}$
is approximated in the following form \citep{Book:Aharonian04}:

\begin{equation}
  \sigma_{pp}(E_\mathrm{kin}) \approx 30 \left[0.95+0.06
  \ln(E_\mathrm{kin}/\mathrm{GeV})\right] \mathrm{mb}
\label{eqA7}
\end{equation}

\noindent The corresponding integrals can be calculated
analytically. Since the spectrum is relatively soft, the resulting
effective cross section $\langle\sigma_\mathrm{pp}\rangle \approx
33$\,mb corresponds to a rather low effective particle energy $\langle
E_\mathrm{kin}\rangle \, \approx 12$\,GeV. Due to the slightly softer
source spectral index compared to the resulting \g-ray spectral index
($s\simeq 2.4$ vs.  $\Gamma_\mathrm{sim} = 2.34$), the assumption of
equal spectral indices infers an error of $\approx 20\%$ of the proton
energy density $E_{pp}^{\pi}$ in Equations~\ref{eq3} and \ref{eq4} of
the main text (S.R. Kelner, priv. comm.). This is certainly a
negligible effect, compared to all other uncertainties.

\bibliographystyle{apj}
\bibliography{ngc253_paper_v3}

\begin{thebibliography}{55}
\expandafter\ifx\csname natexlab\endcsname\relax\def\natexlab#1{#1}\fi

\bibitem[{{Abdo} {et~al.}(2009){Abdo}, {Ackermann}, {Ajello}, {Atwood},
  {Axelsson}, {Baldini}, {Ballet}, {Barbiellini}, {Baring}, {Bastieri},
  {Bechtol}, {Bellazzini}, {Berenji}, {Bloom}, {Bonamente}, {Borgland},
  {Bregeon}, {Brez}, {Brigida}, {Bruel}, {Burnett}, {Caliandro}, {Cameron},
  {Caraveo}, {Casandjian}, {Cavazzuti}, {Cecchi}, {{\c C}elik}, {Chekhtman},
  {Cheung}, {Chiang}, {Ciprini}, {Claus}, {Cohen-Tanugi}, {Cominsky}, {Conrad},
  {Cutini}, {de Angelis}, {de Palma}, {Di Bernardo}, {Silva}, {Drell},
  {Drlica-Wagner}, {Dubois}, {Dumora}, {Farnier}, {Favuzzi}, {Fegan}, {Finke},
  {Focke}, {Fortin}, {Foschini}, {Frailis}, {Fukazawa}, {Funk}, {Fusco},
  {Gargano}, {Gasparrini}, {Gehrels}, {Germani}, {Giavitto}, {Giebels},
  {Giglietto}, {Giommi}, {Giordano}, {Glanzman}, {Godfrey}, {Grenier},
  {Grondin}, {Grove}, {Guillemot}, {Guiriec}, {Hanabata}, {Hayashida}, {Hays},
  {Horan}, {Hughes}, {Jackson}, {J{\'o}hannesson}, {Johnson}, {Johnson},
  {Johnson}, {Kamae}, {Katagiri}, {Kataoka}, {Kawai}, {Kerr}, {Kn{\"o}dlseder},
  {Kocian}, {Kuss}, {Lande}, {Latronico}, {Lemoine-Goumard}, {Longo},
  {Loparco}, {Lott}, {Lovellette}, {Lubrano}, {Madejski}, {Makeev},
  {Mazziotta}, {McConville}, {McEnery}, {Meurer}, {Michelson}, {Mitthumsiri},
  {Mizuno}, {Moiseev}, {Monte}, {Monzani}, {Morselli}, {Moskalenko}, {Murgia},
  {Nolan}, {Norris}, {Nuss}, {Ohsugi}, {Omodei}, {Orlando}, {Ormes}, {Ozaki},
  {Paneque}, {Panetta}, {Parent}, {Pelassa}, {Pepe}, {Pesce-Rollins}, {Piron},
  {Porter}, {Rain{\`o}}, {Rando}, {Razzano}, {Reimer}, {Reimer}, {Reposeur},
  {Reyes}, {Ritz}, {Rochester}, {Rodriguez}, {Roth}, {Ryde}, {Sadrozinski},
  {Sanchez}, {Sander}, {Saz Parkinson}, {Scargle}, {Schalk}, {Sellerholm},
  {Sgr{\`o}}, {Shaw}, {Siskind}, {Smith}, {Smith}, {Spandre}, {Spinelli},
  {Strickman}, {Suson}, {Tajima}, {Takahashi}, {Takahashi}, {Tanaka}, {Tanaka},
  {Thayer}, {Thayer}, {Thompson}, {Tibaldo}, {Torres}, {Tosti}, {Tramacere},
  {Uchiyama}, {Usher}, {Vasileiou}, {Vilchez}, {Vitale}, {Waite}, {Wang},
  {Winer}, {Wood}, {Ylinen}, \& {Ziegler}}]{Fermi:Blazars09}
{Abdo}, A.~A., {et~al.} (Fermi-LAT Collaboration) 2009, \apj, 707, 1310

\bibitem[{{Abdo} {et~al.}(2010{\natexlab{a}}){Abdo}, {Ackermann}, {Ajello},
  {Atwood}, {Axelsson}, {Baldini}, {Ballet}, {Barbiellini}, {Bastieri},
  {Bechtol}, {Bellazzini}, {Berenji}, {Bloom}, {Bonamente}, {Borgland},
  {Bregeon}, {Brez}, {Brigida}, {Bruel}, {Burnett}, {Caliandro}, {Cameron},
  {Caraveo}, {Casandjian}, {Cavazzuti}, {Cecchi}, {{\c C}elik}, {Charles},
  {Chekhtman}, {Cheung}, {Chiang}, {Ciprini}, {Claus}, {Cohen-Tanugi},
  {Conrad}, {Dermer}, {de Angelis}, {de Palma}, {Digel}, {Silva}, {Drell},
  {Drlica-Wagner}, {Dubois}, {Dumora}, {Farnier}, {Favuzzi}, {Fegan}, {Focke},
  {Foschini}, {Frailis}, {Fukazawa}, {Funk}, {Fusco}, {Gargano}, {Gasparrini},
  {Gehrels}, {Germani}, {Giebels}, {Giglietto}, {Giordano}, {Glanzman},
  {Godfrey}, {Grenier}, {Grondin}, {Grove}, {Guillemot}, {Guiriec}, {Hanabata},
  {Harding}, {Hayashida}, {Hays}, {Hughes}, {J{\'o}hannesson}, {Johnson},
  {Johnson}, {Johnson}, {Kamae}, {Katagiri}, {Kataoka}, {Kawai}, {Kerr},
  {Kn{\"o}dlseder}, {Kocian}, {Kuss}, {Lande}, {Latronico}, {Lemoine-Goumard},
  {Longo}, {Loparco}, {Lott}, {Lovellette}, {Lubrano}, {Madejski}, {Makeev},
  {Mazziotta}, {McConville}, {McEnery}, {Meurer}, {Michelson}, {Mitthumsiri},
  {Mizuno}, {Moiseev}, {Monte}, {Monzani}, {Morselli}, {Moskalenko}, {Murgia},
  {Nakamori}, {Nolan}, {Norris}, {Nuss}, {Ohsugi}, {Omodei}, {Orlando},
  {Ormes}, {Ozaki}, {Paneque}, {Panetta}, {Parent}, {Pelassa}, {Pepe},
  {Pesce-Rollins}, {Piron}, {Porter}, {Rain{\`o}}, {Rando}, {Razzano},
  {Reimer}, {Reimer}, {Reposeur}, {Ritz}, {Rodriguez}, {Romani}, {Roth},
  {Ryde}, {Sadrozinski}, {Sander}, {Saz Parkinson}, {Scargle}, {Sellerholm},
  {Sgr{\`o}}, {Shaw}, {Smith}, {Smith}, {Spandre}, {Spinelli}, {Strickman},
  {Strong}, {Suson}, {Takahashi}, {Tanaka}, {Thayer}, {Thayer}, {Thompson},
  {Tibaldo}, {Tibolla}, {Torres}, {Tosti}, {Tramacere}, {Uchiyama}, {Usher},
  {Vasileiou}, {Vilchez}, {Vitale}, {Waite}, {Wang}, {Winer}, {Wood}, {Ylinen},
  \& {Ziegler}}]{Fermi:NGC253M82}
---. 2010{\natexlab{a}}, \apjl, 709, L152

\bibitem[{{Abdo} {et~al.}(2010{\natexlab{b}}){Abdo}, {Ackermann}, {Ajello},
  {Allafort}, {Antolini}, {Atwood}, {Axelsson}, {Baldini}, {Ballet},
  {Barbiellini}, \& et~al.}]{Fermi:1yr}
---. 2010{\natexlab{b}}, \apjs, 188, 405

\bibitem[{{Acciari} {et~al.}(2009){Acciari}, {Aliu}, {Arlen}, \& et al.}]{VERITAS:M82}
{Acciari}, V.~A., {et~al.} (VERITAS Collaboration) 2009, \nat, 462, 770

\bibitem[{{Acero} {et~al.}(2009){Acero}, {Aharonian}, {Akhperjanian}, {Anton},
  {Barres de Almeida}, {Bazer-Bachi}, {Becherini}, {Behera}, {Bernl{\"o}hr},
  {Bochow}, {Boisson}, {Bolmont}, {Borrel}, {Brucker}, {Brun}, {Brun},
  {B{\"u}hler}, {Bulik}, {B{\"u}sching}, {Boutelier}, {Chadwick},
  {Charbonnier}, {Chaves}, {Cheesebrough}, {Chounet}, {Clapson}, {Coignet},
  {Dalton}, {Daniel}, {Davids}, {Degrange}, {Deil}, {Dickinson},
  {Djannati-Ata{\"\i}}, {Domainko}, {Drury}, {Dubois}, {Dubus}, {Dyks},
  {Dyrda}, {Egberts}, {Emmanoulopoulos}, {Espigat}, {Farnier}, {Fegan},
  {Feinstein}, {Fiasson}, {F{\"o}rster}, {Fontaine}, {F{\"u}{\ss}ling},
  {Gabici}, {Gallant}, {G{\'e}rard}, {Gerbig}, {Giebels}, {Glicenstein},
  {Gl{\"u}ck}, {Goret}, {G{\"o}ring}, {Hauser}, {Hauser}, {Heinz},
  {Heinzelmann}, {Henri}, {Hermann}, {Hinton}, {Hoffmann}, {Hofmann},
  {Hofverberg}, {Hoppe}, {Horns}, {Jacholkowska}, {de Jager}, {Jahn}, {Jung},
  {Katarzy{\'n}ski}, {Katz}, {Kaufmann}, {Kerschhaggl}, {Khangulyan},
  {Kh{\'e}lifi}, {Keogh}, {Klochkov}, {Klu{\'z}niak}, {Kneiske}, {Komin},
  {Kosack}, {Kossakowski}, {Lamanna}, {Lenain}, {Lohse}, {Marandon},
  {Martineau-Huynh}, {Marcowith}, {Masbou}, {Maurin}, {McComb}, {Medina},
  {M{\'e}hault}, {Moderski}, {Moulin}, {Naumann-Godo}, {de Naurois}, {Nedbal},
  {Nekrassov}, {Nicholas}, {Niemiec}, {Nolan}, {Ohm}, {Olive}, {Wilhelmi},
  {Orford}, {Ostrowski}, {Panter}, {Arribas}, {Pedaletti}, {Pelletier},
  {Petrucci}, {Pita}, {P{\"u}hlhofer}, {Punch}, {Quirrenbach}, {Raubenheimer},
  {Raue}, {Rayner}, {Reimer}, {Renaud}, {Rieger}, {Ripken}, {Rob},
  {Rosier-Lees}, {Rowell}, {Rudak}, {Rulten}, {Ruppel}, {Sahakian},
  {Santangelo}, {Schlickeiser}, {Sch{\"o}ck}, {Schwanke}, {Schwarzburg},
  {Schwemmer}, {Shalchi}, {Sikora}, {Skilton}, {Sol}, {Stawarz}, {Steenkamp},
  {Stegmann}, {Stinzing}, {Superina}, {Szostek}, {Tam}, {Tavernet}, {Terrier},
  {Tibolla}, {Tluczykont}, {van Eldik}, {Vasileiadis}, {Venter}, {Venter},
  {Vialle}, {Vincent}, {Vivier}, {V{\"o}lk}, {Volpe}, {Wagner}, {Ward},
  {Zdziarski}, \& {Zech}}]{HESS:NGC253}
{Acero}, F., {et~al.} (H.E.S.S. Collaboration) 2009, Science, 326, 1080

%\bibitem[{{Ackermann} {et~al.}(2012),{Ackermann}, {Ajello},
%    {Allafort}, {Baldini}, {Ballet}, {Bastieri}, {Bechtol},
%    {Bellazzini}, {Berenji}, {Bloom}, {Bonamente}, {Borgland},
%    {Bouvier}, {Bregeon}, {Brigida}, {Bruel}, {Buehler}, {Buson},
%    {Caliandro}, {Cameron}, {Caraveo}, {Casandjian}, {Cecchi},
%    {Charles}, {Chekhtman}, {Cheung}, {Chiang}, {Cillis}, {Ciprini},
%    {Claus}, {Cohen-Tanugi}, {Conrad}, {Cutini}, {De Palma}, {Dermer},
%    {Digel}, {Silva}, {Drell}, {Drlica-Wagner}, {Favuzzi}, {Fegan},
%   {Fortin}, {Fukazawa}, {Funk}, {Fusco}, {Gargano}, {Gasparrini},
%    {Germani}, {Giglietto}, {Giordano}, {Glanzman}, {Godfrey},
%    {Grenier}, {Guiriec}, {Gustafsson}, {Hadasch}, {Hayashida},
%    {Hays}, {Hughes}, {J{\'o}hannesson}, {Johnson}, {Kamae},
%    {Katagiri}, {Kataoka}, {Kn{\"o}dlseder}, {Kuss}, {Lande}, {Longo},
%    {Loparco}, {Lott}, {Lovellette}, {Lubrano}, {Madejski}, {Martin},
%    {Mazziotta}, {McEnery}, {Michelson}, {Mizuno}, {Monte}, {Monzani},
%    {Morselli}, {Moskalenko}, {Murgia}, {Nishino}, {Norris}, {Nuss},
%    {Ohno}, {Ohsugi}, {Okumura}, {Omodei}, {Orlando}, {Ozaki},
%    {Parent}, {Persic}, {Pesce-Rollins}, {Petrosian}, {Pierbattista},
%    {Piron}, {Pivato}, {Porter}, {Rain{\`o}}, {Rando}, {Razzano},
%    {Reimer}, {Reimer}, {Ritz}, {Roth}, {Sbarra}, {Sgr{\`o}},
%    {Siskind}, {Spandre}, {Spinelli}, {Stawarz}, {Strong},
%    {Takahashi}, {Tanaka}, {Thayer}, {Tibaldo}, {Tinivella}, {Torres},
%    {Tosti}, {Troja}, {Uchiyama}, {Vandenbroucke}, {Vianello},
%    {Vitale}, {Waite}, {Wood}, \& {Yang}}]{Fermi:SB} {Ackermann}, M.,
%  {et~al.} (Fermi-LAT Collaboration) 2012, ArXiv e-prints:1206.1346

\bibitem[{{Aharonian} {et~al.}(2005){Aharonian}, {Akhperjanian}, {Bazer-Bachi},
  {Beilicke}, {Benbow}, {Berge}, {Bernl{\"o}hr}, {Boisson}, {Bolz}, {Borrel},
  {Braun}, {Breitling}, {Brown}, {Chadwick}, {Chounet}, {Cornils},
  {Costamante}, {Degrange}, {Dickinson}, {Djannati-Ata{\"\i}}, {O'C.~Drury},
  {Dubus}, {Emmanoulopoulos}, {Espigat}, {Feinstein}, {Fontaine}, {Fuchs},
  {Funk}, {Gallant}, {Giebels}, {Gillessen}, {Glicenstein}, {Goret},
  {Hadjichristidis}, {Hauser}, {Heinzelmann}, {Henri}, {Hermann}, {Hinton},
  {Hofmann}, {Holleran}, {Horns}, {Jacholkowska}, {de Jager}, {Kh{\'e}lifi},
  {Komin}, {Konopelko}, {Latham}, {Le Gallou}, {Lemi{\`e}re},
  {Lemoine-Goumard}, {Leroy}, {Lohse}, {Martin}, {Martineau-Huynh},
  {Marcowith}, {Masterson}, {McComb}, {de Naurois}, {Nolan}, {Noutsos},
  {Orford}, {Osborne}, {Ouchrif}, {Panter}, {Pelletier}, {Pita},
  {P{\"u}hlhofer}, {Punch}, {Raubenheimer}, {Raue}, {Raux}, {Rayner}, {Reimer},
  {Reimer}, {Ripken}, {Rob}, {Rolland}, {Rowell}, {Sahakian}, {Saug{\'e}},
  {Schlenker}, {Schlickeiser}, {Schuster}, {Schwanke}, {Siewert}, {Sol},
  {Spangler}, {Steenkamp}, {Stegmann}, {Tavernet}, {Terrier}, {Th{\'e}oret},
  {Tluczykont}, {Vasileiadis}, {Venter}, {Vincent}, {V{\"o}lk}, \&
  {Wagner}}]{HESS:NGC253_UL}
{Aharonian}, F., {et~al.} (H.E.S.S. Collaboration) 2005, \aap, 442, 177

\bibitem[{{Aharonian} {et~al.}(2006){Aharonian}, {Akhperjanian}, {Bazer-Bachi},
  {Beilicke}, {Benbow}, {Berge}, {Bernl{\"o}hr}, {Boisson}, {Bolz}, {Borrel},
  {Braun}, {Breitling}, {Brown}, {B{\"u}hler}, {B{\"u}sching}, {Carrigan},
  {Chadwick}, {Chounet}, {Cornils}, {Costamante}, {Degrange}, {Dickinson},
  {Djannati-Ata{\"\i}}, {O'C.~Drury}, {Dubus}, {Egberts}, {Emmanoulopoulos},
  {Espigat}, {Feinstein}, {Ferrero}, {Fiasson}, {Fontaine}, {Funk}, {Funk},
  {Gallant}, {Giebels}, {Glicenstein}, {Goret}, {Hadjichristidis}, {Hauser},
  {Hauser}, {Heinzelmann}, {Henri}, {Hermann}, {Hinton}, {Hofmann}, {Holleran},
  {Horns}, {Jacholkowska}, {de Jager}, {Kh{\'e}lifi}, {Komin}, {Konopelko},
  {Kosack}, {Latham}, {Le Gallou}, {Lemi{\`e}re}, {Lemoine-Goumard}, {Lohse},
  {Martin}, {Martineau-Huynh}, {Marcowith}, {Masterson}, {McComb}, {de
  Naurois}, {Nedbal}, {Nolan}, {Noutsos}, {Orford}, {Osborne}, {Ouchrif},
  {Panter}, {Pelletier}, {Pita}, {P{\"u}hlhofer}, {Punch}, {Raubenheimer},
  {Raue}, {Rayner}, {Reimer}, {Reimer}, {Ripken}, {Rob}, {Rolland}, {Rowell},
  {Sahakian}, {Saug{\'e}}, {Schlenker}, {Schlickeiser}, {Schwanke}, {Sol},
  {Spangler}, {Spanier}, {Steenkamp}, {Stegmann}, {Superina}, {Tavernet},
  {Terrier}, {Th{\'e}oret}, {Tluczykont}, {van Eldik}, {Vasileiadis}, {Venter},
  {Vincent}, {V{\"o}lk}, {Wagner}, \& {Ward}}]{HESS:Crab}
---. 2006, \aap, 457, 899

\bibitem[{{Aharonian}(2004)}]{Book:Aharonian04}
{Aharonian}, F.~A. 2004, {Very high energy cosmic gamma radiation : a crucial
  window on the extreme Universe} (World Scientific Publishing Company; 1st
  edition (April 30, 2003))

\bibitem[{{Akyuz} {et~al.}(1991){Akyuz}, {Brouillet}, \& {Ozel}}]{M82:Akyuz91}
{Akyuz}, A., {Brouillet}, N., \& {Ozel}, M.~E. 1991, \aap, 248, 419

\bibitem[{{Atwood} {et~al.}(2009){Atwood}, {Abdo}, {Ackermann}, {Althouse},
  {Anderson}, {Axelsson}, {Baldini}, {Ballet}, {Band}, {Barbiellini}, \&
  et~al.}]{Fermi:Inst}
{Atwood}, W.~B., {et~al.} (Fermi-LAT Collaboration) 2009, \apj, 697, 1071

\bibitem[{{Berezinskii} {et~al.}(1990){Berezinskii}, {Bulanov}, {Dogiel}, \&
  {Ptuskin}}]{Berezinskii90}
{Berezinskii}, V.~S., {Bulanov}, S.~V., {Dogiel}, V.~A., \& {Ptuskin}, V.~S.
  1990, {Astrophysics of cosmic rays}, ed. {Berezinskii, V.~S., Bulanov, S.~V.,
  Dogiel, V.~A., \&amp; Ptuskin, V.~S. } (Elsevier Science Ltd)

\bibitem[{{Berge} {et~al.}(2007){Berge}, {Funk}, \& {Hinton}}]{HESS:Background}
{Berge}, D., {Funk}, S., \& {Hinton}, J. 2007, \aap, 466, 1219

\bibitem[{{Blandford} \& {Eichler}(1987)}]{Blandford87}
{Blandford}, R., \& {Eichler}, D. 1987, \physrep, 154, 1

\bibitem[{{Breitschwerdt} {et~al.}(1991){Breitschwerdt}, {McKenzie}, \&
  {V\"olk}}]{Breitschwerdt91}
{Breitschwerdt}, D., {McKenzie}, J.~F., \& {V\"olk}, H.~J. 1991, \aap, 245, 79

\bibitem[{{Bykov}(2001)}]{Bykov01}
{Bykov}, A.~M. 2001, Space Science Reviews, 99, 317

\bibitem[{{Dalcanton} {et~al.}(2009){Dalcanton}, {Williams}, {Seth}, {Dolphin},
  {Holtzman}, {Rosema}, {Skillman}, {Cole}, {Girardi}, {Gogarten},
  {Karachentsev}, {Olsen}, {Weisz}, {Christensen}, {Freeman}, {Gilbert},
  {Gallart}, {Harris}, {Hodge}, {de Jong}, {Karachentseva}, {Mateo}, {Stetson},
  {Tavarez}, {Zaritsky}, {Governato}, \& {Quinn}}]{Dalcanton09}
{Dalcanton}, J.~J., {et~al.} 2009, \apjs, 183, 67

\bibitem[{{Davidge} {et~al.}(1991){Davidge}, {Le Fevre}, \&
  {Clark}}]{NGC253:Davidge91}
{Davidge}, T.~J., {Le Fevre}, O., \& {Clark}, C.~C. 1991, \apj, 370, 559

\bibitem[{{de Naurois} \& {Rolland}(2009)}]{Model++}
{de Naurois}, M., \& {Rolland}, L. 2009, Astroparticle Physics, 32, 231

\bibitem[{{Domingo-Santamar{\'{\i}}a} \& {Torres}(2005)}]{NGC253:Domingo05}
{Domingo-Santamar{\'{\i}}a}, E., \& {Torres}, D.~F. 2005, \aap, 444, 403

\bibitem[{{Drury} {et~al.}(1989){Drury}, {Markiewicz}, \& {V\"olk}}]{Drury89}
{Drury}, L.~O., {Markiewicz}, W.~J., \& {V\"olk}, H.~J. 1989, \aap, 225, 179

\bibitem[{{Engelbracht} {et~al.}(1998){Engelbracht}, {Rieke}, {Rieke}, {Kelly},
  \& {Achtermann}}]{NGC253:Engelbracht98}
{Engelbracht}, C.~W., {Rieke}, M.~J., {Rieke}, G.~H., {Kelly}, D.~M., \&
  {Achtermann}, J.~M. 1998, \apj, 505, 639

\bibitem[{{Feldman} \& {Cousins}(1998)}]{FeldmanCousins98}
{Feldman}, G.~J., \& {Cousins}, R.~D. 1998, {\prd}, 57, 3873

\bibitem[{{Heesen} {et~al.}(2009){Heesen}, {Beck}, {Krause}, \&
  {Dettmar}}]{NGC253:Heesen09}
{Heesen}, V., {Beck}, R., {Krause}, M., \& {Dettmar}, R.-J. 2009, \aap, 494,
  563

\bibitem[{{Heesen} {et~al.}(2011){Heesen}, {Beck}, {Krause}, \&
  {Dettmar}}]{Heesen11}
---. 2011, \aap, 535, A79

\bibitem[{{Hillas}(1985)}]{Hillas85}
{Hillas}, A.~M. 1985, in International Cosmic Ray Conference, ed.
  {F.~C.~Jones}, Vol.~3, {445--448}

\bibitem[{{Inoue}(2011)}]{Inoue11}
{Inoue}, Y. 2011, \apj, 728, 11

\bibitem[{{Kamae} {et~al.}(2006){Kamae}, {Karlsson}, {Mizuno}, {Abe}, \&
  {Koi}}]{Kamae2006}
{Kamae}, T., {Karlsson}, N., {Mizuno}, T., {Abe}, T., \& {Koi}, T. 2006, \apj,
  647, 692

\bibitem[{{Karachentsev} {et~al.}(2003){Karachentsev}, {Grebel}, {Sharina},
  {Dolphin}, {Geisler}, {Guhathakurta}, {Hodge}, {Karachentseva}, {Sarajedini},
  \& {Seitzer}}]{NGC253:Kara03}
{Karachentsev}, I.~D., {et~al.} 2003, \aap, 404, 93

\bibitem[{{Karlsson}(2008)}]{Karlsson2008}
{Karlsson}, N. 2008, in American Institute of Physics Conference Series, Vol.
  1085, American Institute of Physics Conference Series, ed. F.~A. {Aharonian},
  W.~{Hofmann}, \& F.~{Rieger}, 561--564

\bibitem[{{Kelner} {et~al.}(2006){Kelner}, {Aharonian}, \&
  {Bugayov}}]{Kelner2006}
{Kelner}, S.~R., {Aharonian}, F.~A., \& {Bugayov}, V.~V. 2006, \prd, 74, 034018

\bibitem[{{Lacki} {et~al.}(2010){Lacki}, {Thompson}, \&
  {Quataert}}]{NGC253:Lacki10}
{Lacki}, B.~C., {Thompson}, T.~A., \& {Quataert}, E. 2010, \apj, 717, 1

\bibitem[{{Lacki} {et~al.}(2011){Lacki}, {Thompson}, {Quataert}, {Loeb}, \&
  {Waxman}}]{NGC253:Lacki11}
{Lacki}, B.~C., {Thompson}, T.~A., {Quataert}, E., {Loeb}, A., \& {Waxman}, E.
  2011, \apj, 734, 107

%\bibitem[{{Lenain} {et~al.}(2010){Lenain}, {Ricci}, {T{\"u}rler}, {Dorner}, \&
%  {Walter}}]{Lenain2010}
%{Lenain}, J.-P., {Ricci}, C., {T{\"u}rler}, M., {Dorner}, D., \& {Walter}, R.
%  2010, \aap, 524, A72

\bibitem[{{Li} \& {Ma}(1983)}]{LiMa}
{Li}, T., \& {Ma}, Y. 1983, \apj, 272, 317

\bibitem[{{Loeb} \& {Waxman}(2006)}]{Loeb06}
{Loeb}, A., \& {Waxman}, E. 2006, \jcap, 5, 3

\bibitem[{{Mannheim} {et~al.}(2012){Mannheim}, {Els{\"a}sser}, \&
  {Tibolla}}]{NGC253:Mannheim12}
{Mannheim}, K., {Els{\"a}sser}, D., \& {Tibolla}, O. 2012, Astroparticle
  Physics, 35, 797

\bibitem[{{Mattox} {et~al.}(1996){Mattox}, {Bertsch}, {Chiang}, {Dingus},
  {Digel}, {Esposito}, {Fierro}, {Hartman}, {Hunter}, {Kanbach}, {Kniffen},
  {Lin}, {Macomb}, {Mayer-Hasselwander}, {Michelson}, {von Montigny},
  {Mukherjee}, {Nolan}, {Ramanamurthy}, {Schneid}, {Sreekumar}, {Thompson}, \&
  {Willis}}]{Mattox1996}
{Mattox}, J.~R., {et~al.} 1996, \apj, 461, 396

\bibitem[{{Mauersberger} {et~al.}(1996){Mauersberger}, {Henkel}, {Wielebinski},
  {Wiklind}, \& {Reuter}}]{NGC253:Mauersberger96}
{Mauersberger}, R., {Henkel}, C., {Wielebinski}, R., {Wiklind}, T., \&
  {Reuter}, H.-P. 1996, \aap, 305, 421

\bibitem[{{Melo} {et~al.}(2002){Melo}, {P{\'e}rez Garc{\'{\i}}a},
  {Acosta-Pulido}, {Mu{\~n}oz-Tu{\~n}{\'o}n}, \& {Rodr{\'{\i}}guez
  Espinosa}}]{NGC253:Melo}
{Melo}, V.~P., {P{\'e}rez Garc{\'{\i}}a}, A.~M., {Acosta-Pulido}, J.~A.,
  {Mu{\~n}oz-Tu{\~n}{\'o}n}, C., \& {Rodr{\'{\i}}guez Espinosa}, J.~M. 2002,
  \apj, 574, 709

\bibitem[{{Musulmanbekov}(2004)}]{PPinelas03}
{Musulmanbekov}, G. 2004, Physics of Atomic Nuclei, 67, 90

\bibitem[{{Nolan} {et~al.}(2012){Nolan}, {Abdo}, {Ackermann}, {Ajello},
  {Allafort}, {Antolini}, {Atwood}, {Axelsson}, {Baldini}, {Ballet}, \&
  et~al.}]{Fermi:2yr}
{Nolan}, P.~L., {et~al.} (Fermi-LAT Collaboration) 2012, \apjs, 199, 31

\bibitem[{{Ohm} \& {Hinton}(2012)}]{OhmHinton12}
{Ohm}, S., \& {Hinton}, J.~A. 2012, ArXiv e-prints: 1202.0260

\bibitem[{{Ohm} {et~al.}(2009){Ohm}, {van Eldik}, \& {Egberts}}]{TMVA}
{Ohm}, S., {van Eldik}, C., \& {Egberts}, K. 2009, Astroparticle Physics, 31,
  383

\bibitem[{{Paglione} {et~al.}(1996){Paglione}, {Marscher}, {Jackson}, \&
  {Bertsch}}]{NGC253:Paglione96}
{Paglione}, T.~A.~D., {Marscher}, A.~P., {Jackson}, J.~M., \& {Bertsch}, D.~L.
  1996, \apj, 460, 295

\bibitem[{{Paglione} {et~al.}(1995){Paglione}, {Tosaki}, \&
  {Jackson}}]{NGC253:Paglione95}
{Paglione}, T.~A.~D., {Tosaki}, T., \& {Jackson}, J.~M. 1995, \apjl, 454, L117+

\bibitem[{{Parizot} {et~al.}(2004){Parizot}, {Marcowith}, {van der Swaluw},
  {Bykov}, \& {Tatischeff}}]{Parizot04}
{Parizot}, E., {Marcowith}, A., {van der Swaluw}, E., {Bykov}, A.~M., \&
  {Tatischeff}, V. 2004, \aap, 424, 747

\bibitem[{{Pence}(1980)}]{NGC253:Pence80}
{Pence}, W.~D. 1980, \apj, 239, 54

\bibitem[{{Ptuskin} {et~al.}(1997){Ptuskin}, {V\"olk}, {Zirakashvili}, \&
  {Breitschwerdt}}]{Ptuskin97}
{Ptuskin}, V.~S., {V\"olk}, H.~J., {Zirakashvili}, V.~N., \& {Breitschwerdt},
  D. 1997, \aap, 321, 434

\bibitem[{{Rekola} {et~al.}(2005){Rekola}, {Richer}, {McCall}, {Valtonen},
  {Kotilainen}, \& {Flynn}}]{NGC253:Rekola05}
{Rekola}, R., {Richer}, M.~G., {McCall}, M.~L., {Valtonen}, M.~J.,
  {Kotilainen}, J.~K., \& {Flynn}, C. 2005, \mnras, 361, 330

\bibitem[{{Rephaeli} {et~al.}(2010){Rephaeli}, {Arieli}, \&
  {Persic}}]{NGC253:Rephaeli10}
{Rephaeli}, Y., {Arieli}, Y., \& {Persic}, M. 2010, \mnras, 401, 473

\bibitem[{{Sakamoto} {et~al.}(2011){Sakamoto}, {Mao}, {Matsushita}, {Peck},
  {Sawada}, \& {Wiedner}}]{NGC253:Sakamoto11}
{Sakamoto}, K., {Mao}, R.-Q., {Matsushita}, S., {Peck}, A.~B., {Sawada}, T., \&
  {Wiedner}, M.~C. 2011, \apj, 735, 19

\bibitem[{{Sorai} {et~al.}(2000){Sorai}, {Nakai}, {Kuno}, {Nishiyama}, \&
  {Hasegawa}}]{NGC253:Sorai00}
{Sorai}, K., {Nakai}, N., {Kuno}, N., {Nishiyama}, K., \& {Hasegawa}, T. 2000,
  \pasj, 52, 785

\bibitem[{{Strong} {et~al.}(2007){Strong}, {Moskalenko}, \&
  {Ptuskin}}]{Strong2007}
{Strong}, A.~W., {Moskalenko}, I.~V., \& {Ptuskin}, V.~S. 2007, Annual Review
  of Nuclear and Particle Science, 57, 285

\bibitem[{{Strong} {et~al.}(2010){Strong}, {Porter}, {Digel},
  {J{\'o}hannesson}, {Martin}, {Moskalenko}, {Murphy}, \&
  {Orlando}}]{MW:Strong10}
{Strong}, A.~W., {Porter}, T.~A., {Digel}, S.~W., {J{\'o}hannesson}, G.,
  {Martin}, P., {Moskalenko}, I.~V., {Murphy}, E.~J., \& {Orlando}, E. 2010,
  \apjl, 722, L58

\bibitem[{{Thompson} {et~al.}(2007){Thompson}, {Quataert}, \&
  {Waxman}}]{Thompson07}
{Thompson}, T.~A., {Quataert}, E., \& {Waxman}, E. 2007, \apj, 654, 219

\bibitem[{{Van Buren} \& {Greenhouse}(1994)}]{VanBuren94}
{Van Buren}, D., \& {Greenhouse}, M.~A. 1994, \apj, 431, 640

\bibitem[{{V\"olk} {et~al.}(1996){V\"olk}, {Aharonian}, \&
  {Breitschwerdt}}]{Voelk96}
{V\"olk}, H.~J., {Aharonian}, F.~A., \& {Breitschwerdt}, D. 1996, \ssr, 75, 279

\bibitem[{{V\"olk} {et~al.}(1989){V\"olk}, {Klein}, \& {Wielebinski}}]{Voelk89}
{V\"olk}, H.~J., {Klein}, U., \& {Wielebinski}, R. 1989, \aap, 213, L12

\bibitem[{{Weaver} {et~al.}(2002){Weaver}, {Heckman}, {Strickland}, \&
  {Dahlem}}]{NGC253:Weaver02}
{Weaver}, K.~A., {Heckman}, T.~M., {Strickland}, D.~K., \& {Dahlem}, M. 2002,
  \apjl, 576, L19

\bibitem[{{Zirakashvili} \& {V{\"o}lk}(2006)}]{NGC253:Zirakashvili06}
{Zirakashvili}, V.~N., \& {V{\"o}lk}, H.~J. 2006, \apj, 636, 140

\end{thebibliography}
\end{document}